\documentclass[aps,prl,reprint,superscriptaddress]{revtex4-2}

\usepackage{amsmath,amssymb}
\usepackage[draft,defaultcolor=red,commandnameprefix=ifneeded,authormarkup=none]{changes}
\definechangesauthor[name={Praful}, color=Rhodamine]{PG}
\definechangesauthor[name={David}, color=Greene]{DL}
\definechangesauthor[name={Jiwei}, color=Blue]{JW}
\definechangesauthor[name={Yannick}, color=Red]{YD}

\usepackage{bm}
\usepackage{graphicx}
\usepackage[hidelinks]{hyperref}
\usepackage{comment}
\usepackage{subcaption} 
\usepackage[dvipsnames]{xcolor}
\usepackage{enumitem} 
\usepackage[normalem]{ulem}
\usepackage{bibunits}
\defaultbibliography{bibliography}      
\defaultbibliographystyle{apsrev4-2}  

\newcommand{\DL}[1]{{\color{Green}#1 (David)} }
\newcommand{\JW}[1]{{\color{Blue}#1 (Jiwei)} }

\newcommand{\be}{\begin{equation}}
\newcommand{\ee}{\end{equation}}
\newcommand{\bea}{\begin{eqnarray}}
\newcommand{\eea}{\end{eqnarray}}
\newcommand{\ba}{\begin{align}}
\newcommand{\ea}{\end{align}}
\newcommand{\nn}{\nonumber}

\newcommand\mbf[1]{\mathbf #1}
\newcommand\mbb[1]{\mathbb #1}
\newcommand\mcl[1]{\mathcal #1}

\def\St{\mathbb{S}}
\def\Sp{\mathbb{S}^+}
\def\Sm{\mathbb{S}^-}
\def\rSp{\underline{\mathbb{S}}^+}
\def\rSm{\underline{\mathbb{S}}^-}

\def\n{\mathbf{n}}
\def\J{\mathbf{J}}
\def\x{\mathbf{x}}

\def\rx{\underline{\x}}
\def\q{\mathbf{q}}
\def\0{\mathbf{0}}
\def\w{\mathbf{w}}
\newcommand{\jj}{\boldsymbol{\jmath}}

\begin{document}


\title{Topological bounds on the dynamical growth rate of chemical reaction networks}

\author{Praful Gagrani}
\affiliation{Institute of Industrial Science, The University of Tokyo, Tokyo 153-8505, Japan}
\affiliation{Theoretical Sciences Visiting Program, Okinawa Institute of Science and Technology Graduate University, Onna, 904-0495, Japan
}

\author{Jiwei Wang}
\affiliation{Gulliver Laboratory, UMR CNRS 7083, Paris Sciences et Lettres University, Paris F-75231, France}

\author{Yannick De Decker}
\affiliation{Center for Nonlinear Phenomena and Complex Systems (CENOLI), Universit\'e libre de Bruxelles (ULB), Campus Plaine, C.P.\ 231, B-1050 Brussels, Belgium}

\author{David Lacoste}
\affiliation{Theoretical Sciences Visiting Program, Okinawa Institute of Science and Technology Graduate University, Onna, 904-0495, Japan
}
\affiliation{Gulliver Laboratory, UMR CNRS 7083, Paris Sciences et Lettres University, Paris F-75231, France}
\date{\today}

\begin{abstract}
Growth and decay are system-level properties of chemical reaction networks (CRNs) relevant from prebiotic chemistry to cellular metabolism. Their properties are typically analyzed through the kinetics of particular models, which requires specification of the full set of kinetic laws and parameters. In this work, assuming a steady balanced-growth regime, we derive stoichiometry-based constraints on the growth (or shrinkage) rate. The resulting bounds are controlled by a topological quantity, the maximum amplification factor, defined via a von Neumann max–min problem over feasible fluxes as illustrated by numerical tests on random-network ensembles of CRNs. We argue for the relevance of our results in the context of origins of life studies and the design of synthetic chemical reaction networks. 
\end{abstract}

\maketitle
\begin{bibunit}

Autocatalysis and positive feedback are ubiquitous in biochemical organization and in candidate prebiotic chemistries: they enable self-maintenance, replication-like amplification, and, in open settings, persistent growth 
\cite{eigen1977principle,baum2023ecology,ameta_self-reproduction_2021}. Stoichiometric frameworks have recently made the identification of autocatalytic subnetworks operational via purely structural criteria and algorithms that identify minimal autocatalytic motifs and their embeddings inside large networks \cite{blokhuis2020universal,gagrani2024polyhedral,golnik2025enumeration,andersen_defining_2020}. This progress raises a natural quantitative question: once a network is topologically capable of amplification, what controls \emph{how fast} it can grow or decay?

From a dynamical perspective, growth rates depend on nonlinear kinetics and on rate constants that can vary over orders of magnitude and are often poorly constrained. This creates a gap: topology-based notions of autocatalysis certify feasibility, whereas kinetics determines realized performance.
Here we identify a broad balanced-growth regime in which a topological quantity, together with a kinetic prefactor, controls the achievable growth rate. Since this topological quantity depends only on stoichiometric data, which the same type of data used in Flux Balance Analysis (FBA)~\cite{fang_reconstructing_2020}, it provides a novel network-level analytic tool when detailed kinetic information is unavailable.

\textit{Balanced-growth regime---}
A CRN with internal species $\mcl{S}$ and reactions $\mcl{R}$ is specified by nonnegative output and input matrices $(\Sp,\Sm)$. Its stoichiometric matrix is
\(
\St=\Sp-\Sm .
\)
We do not assume reversibility: each column of $\St$ denotes a distinct one-way reaction. The dynamics are
\begin{equation}
\label{eq:rateeq_main}
\dot{\n}=\St\,\J(\n),
\end{equation}
where $\n$ collects species numbers or abundances and $\J(\n)$ the one-way reaction fluxes. Thus topology is encoded in $\St$, while kinetics enter through $\J$.

Given the vector of abundances $\n$, we introduce the extensive scale $\Omega= \w^T \n$ where the vector $\w$ contains the individual scale contribution of each species. For example, if $\n$ is measured in moles and $\Omega$ is the volume, $\w$ is the vector of species-wise molar volumes.
We also define the composition vector $\q = \n/\Omega$, such that 
$\w^T \q = 1.$ For the choice
$\w=[1,\ldots,1]^T$, the components of $\q$ are the relative proportions
of the species in the system. 


The dynamics induced on the composition coordinates are then (SM)
\begin{equation}
\label{eq:q_dynamics_main}
\dot{\q}
=\St\,\frac{\J(\n)}{\Omega}-\frac{ \w^{T}\St\,\J(\n)}{\Omega} \q.
\end{equation}
The fluxes are extensive if they satisfy the property that  $\J(\n)/\Omega = \jj(\q)$ depends only on intensive composition variables $\q$. Under extensivity of the scale and fluxes, a fixed point $\q^*$ of Eq.~\eqref{eq:q_dynamics_main} satisfies
\begin{equation}
\label{eq:LambdaDef}
\St\, \jj(\q^*)=\Lambda\,\q^*,
\qquad
\Lambda=\w^{T}\St\, \jj(\q^*)
\end{equation}
and $\dot{\Omega}=\Lambda\,\Omega$ (SM).
Thus, the existence of this fixed point leads to balanced growth in which all species grow exponentially at common rate $\Lambda$, while their relative populations remain fixed~\cite{lin2020origin,singh2023multistable,dourado_analytical_2020}. We call systems with $\Lambda>0$, $\Lambda=0$, and $\Lambda<0$ growing, steady, and shrinking, respectively.

In this work, we use the linear-volume mass-action form
\begin{align}
J_r(\n)
    =
    k_r \Omega
    \prod_{s \in \mcl{S}}
    \left( \frac{n_s}{\Omega} \right)^{(\Sm)^r_s},
    \label{eq:mass-action_prop}
\end{align}
with rate constants \(k_r>0\). This flux is homogeneous of degree one in
\(\n\)~\cite{pandey2020exponential}. The class of degree-one homogeneous rate laws
with linear volume is broader than mass-action kinetics and includes, for
example, Michaelis--Menten and generalized mass-action forms
\cite{lin2020origin,Liebermeister2010}. 
Eq.~\eqref{eq:mass-action_prop} describes ideal nondilute well-mixed systems, and, in the SM, we show that the usual dilute constant-volume mass-action form is recovered as a limiting case. This choice is therefore natural for
models of cellular growth~\cite{pandey2016analytic,muller2022elementary}.

\textit{Maximum Amplification Factor---}
Independently of kinetics, we consider nonnegative flux vectors $\x\ge 0$ as arbitrary reaction fluxes. Given $\x$, the induced species-wise input and output rates are $\Sm\x$ and $\Sp\x$, respectively (SM). The topological quantity introduced next is defined by optimizing over such fluxes.

 We define the \textit{Amplification Factor} (AF) of a flux vector $\x$ as the species-wise minimum ratio of output to input rates
 \begin{equation}
\label{eq:AF}
\alpha(\x) \;=\; \;\min_{s\in\mathcal{S}}
\frac{(\Sp\x)_s}{(\Sm\x)_s}.
\end{equation}
The \textit{Maximum Amplification Factor} (MAF) is the maximal value of the amplification factor over feasible fluxes,
\begin{equation}
\label{eq:MAF}
\alpha(\Sp,\Sm) \,=\, \max_{\x>\bm{0}} \, \alpha(\x)
\,=\,\max_{\x>\bm{0}}\;\min_{s\in\mathcal{S}}
\frac{(\Sp\x)_s}{(\Sm\x)_s},
\end{equation}
with the convention that species with \((\Sm\x)_s=0\) do not constrain the minimum (they correspond to \(+\infty\)). Note that $\x>\0$ denotes that $\x\geq \0$ but $\x\neq \0$. 

Inspired by von Neumann's technological expansion rate in input-output models of economics (\cite{neumann1945model,sargent2019neumann}, Sec.~9.5 in \cite{gale1989theory}), the MAF was introduced for CRNs by one of the authors 
in \cite{blanco2025identifyingselfamplifyinghypergraphstructures}. Analogous to how the slowest step in a multistep CRN sets the rate-determining step, the MAF is set by the limiting species. By maximizing the smallest species-wise production-to-consumption ratio, the MAF measures the largest growth factor achievable by the network as a whole.
We denote the MAF by $\alpha$ and its optimal flux vector by $\x_\alpha$ henceforth, for clarity.

\textit{Example---} Consider the simplest autocatalytic CRN with two reactions(type I):
\be
    r_1: \text{A} \to \text{B}, \qquad r_2: \text{B} \to 2\text{A}. \nonumber
\ee
Then (ordering species as (A,B) and reactions as ($r_1,r_2$)), 
 \begin{align}
     \Sp &= 
     \begin{bmatrix}
0 & 2 \\
1 & 0
\end{bmatrix},
& 
     \Sm &= 
     \begin{bmatrix}
1 & 0 \\
0 & 1
\end{bmatrix},
&
     \St &= 
     \begin{bmatrix}
-1 & 2 \\
1 & -1
\end{bmatrix}.
 \end{align}
 Following Eq.~\ref{eq:MAF}, 
 \be
\alpha(\Sp,\Sm) = 
\max_{\x>\0} \min\left(\frac{2 \,x_2}{x_1}, \frac{x_1}{x_2}\right). 
 \ee
For any $x_2$, the max-min is attained at $x_1 = \sqrt{2} \, x_2$. Thus,
\bea
\alpha(\Sp,\Sm) &=\sqrt{2}, & \; \; \x_\alpha = [\sqrt{2},1]^\top.
\eea

\textit{Autonomous and autocatalytic networks---} We say that a CRN is \textit{autonomous} if every reaction has at least one reactant and at least one product and every species appears as a reactant in at least one reaction and as a product in at least one reaction. In practice, reactions that effectively create or remove matter are modeled through chemostatted species representing exchange with an environment. Given an arbitrary network, an autonomous subnetwork can be obtained by iteratively removing chemostatted species and species that appear only as reactants or only as products, together with reactions that
act as a source or sink. Thus, autonomous networks are natural candidates for the self-maintaining biochemical core of a cell, as they describe the internally closed reaction architecture required to sustain ongoing biochemical activity. 

For autonomous networks, it can be shown that the MAF satisfies \(0 < \alpha < \infty\) (SM). A CRN is \textit{autocatalytic} if, in addition to being autonomous, there exists a reaction flux such that all species are net produced (SM). From Equation~\ref{eq:MAF}, it follows immediately that $\alpha > 1$ for autocatalytic networks. In SM, we show that the MAF of reducible networks is determined by their irreducible component with the highest MAF; thus, \(\alpha\) acts as a “best-module” topological score. For a given network, efficient algorithms and their implementation for identifying irreducible autocatalytic subnetworks with the highest MAF can be found in \cite{blanco2025identifyingselfamplifyinghypergraphstructures}.

A network is considered to be minimal with respect to a certain property if it does not contain any subnetwork that also verify the property. In particular, finding the MAF for minimal autonomous networks reduces to solving the generalized-eigenvalue problem of the form \(\Sp \x = \alpha \phantom{\cdot}\Sm \x.\) Minimal autocatalytic networks, termed (autocatalytic) \textit{cores}, in reversible networks are shown to be of five distinct types \cite{blokhuis2020universal}, displayed in  Fig.~\ref{fig:coreFig}. These cores provide useful reference points for interpreting \(\alpha\) as an amplification strength (see  Table~\ref{tab:MAFcores}). Since for minimal networks, Eq.~\ref{eq:MAF} reduces to the generalized eigenvalue problem, the `dual' autoinhibitory core resulting from reaction reversal exchanges \(\Sp\leftrightarrow\Sm\) has MAF \(1/\alpha\). Note that, however, for an arbitrary network, \(\alpha(\Sm,\Sp) \neq 1/\alpha(\Sp,\Sm)\).  
\begin{figure}[t]
    \centering 
    \includegraphics[width=0.43\textwidth]{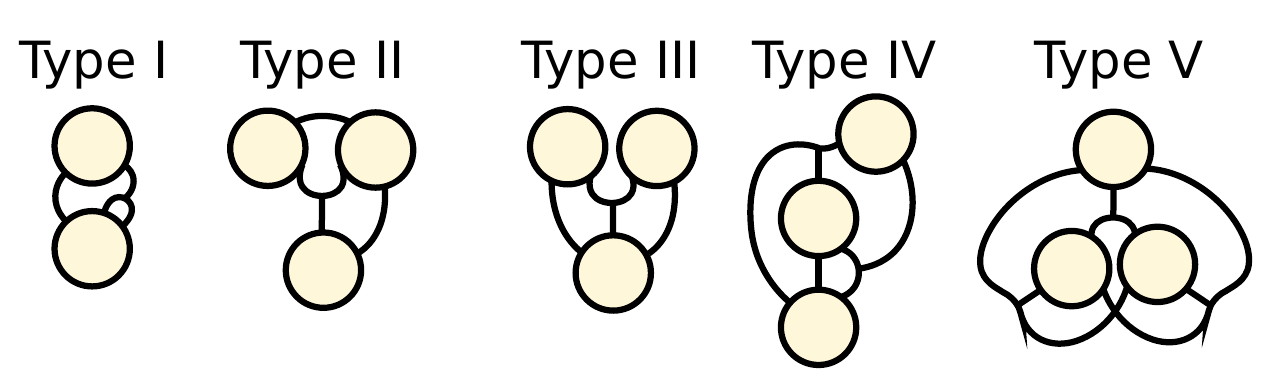}
    \caption{Five types of minimal autocatalytic cores (from Ref.~\cite{blokhuis2020universal}). Reversing all reactions exchanges \(\Sp\leftrightarrow\Sm\) and maps the MAF as \(\alpha\mapsto 1/\alpha\), yielding the corresponding dual autoinhibitory motif.}
    \label{fig:coreFig}
\end{figure}

\begin{table}[t!]
\centering
\begin{tabular}{c c c c c c}
\hline
 & Type I & Type II & Type III & Type IV & Type V \\
\hline
\(\alpha\) (core) & $\sqrt{2}$ & $\approx 1.32$ & $\sqrt{2}$ & $(\sqrt{5}-1)/2$  & 2 \\
\(\alpha\) (dual) & $1/\sqrt{2}$ & $\approx 0.75$ & $1/\sqrt{2}$ & $(\sqrt{5}+1)/2$ & $1/2$ \\
\hline
\end{tabular}
\caption{Representative MAF values for autocatalytic cores and their autoinhibitory duals, consistent with \cite{blokhuis2020universal}.}
\label{tab:MAFcores}
\end{table}

\begin{figure*}[t]
    \centering
    
    \begin{subfigure}{0.44\textwidth}
        \centering
        \includegraphics[width=\linewidth]{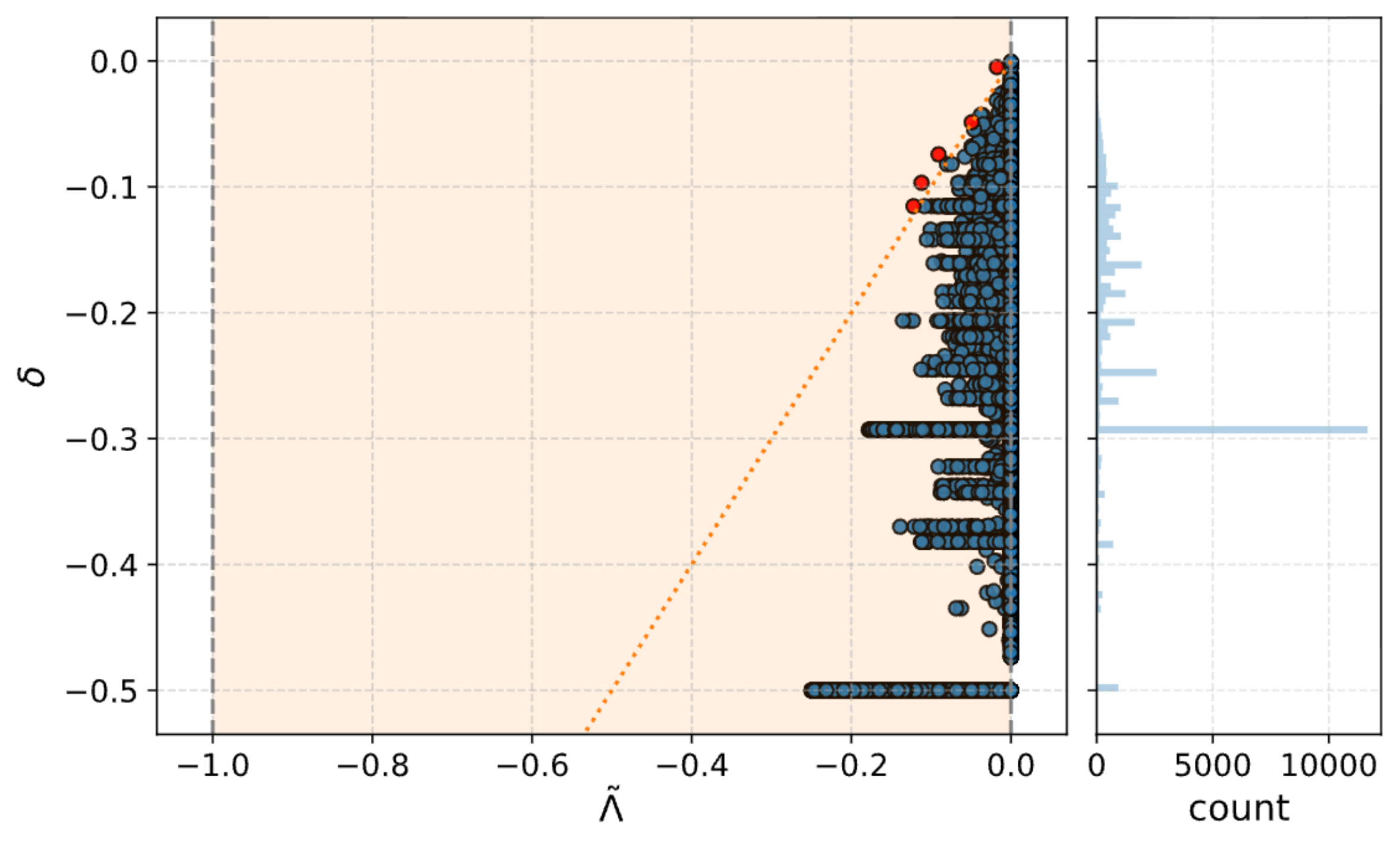}
    \end{subfigure}
    \hfill
    \begin{subfigure}{0.44\textwidth}
        \centering
        \includegraphics[width=\linewidth]{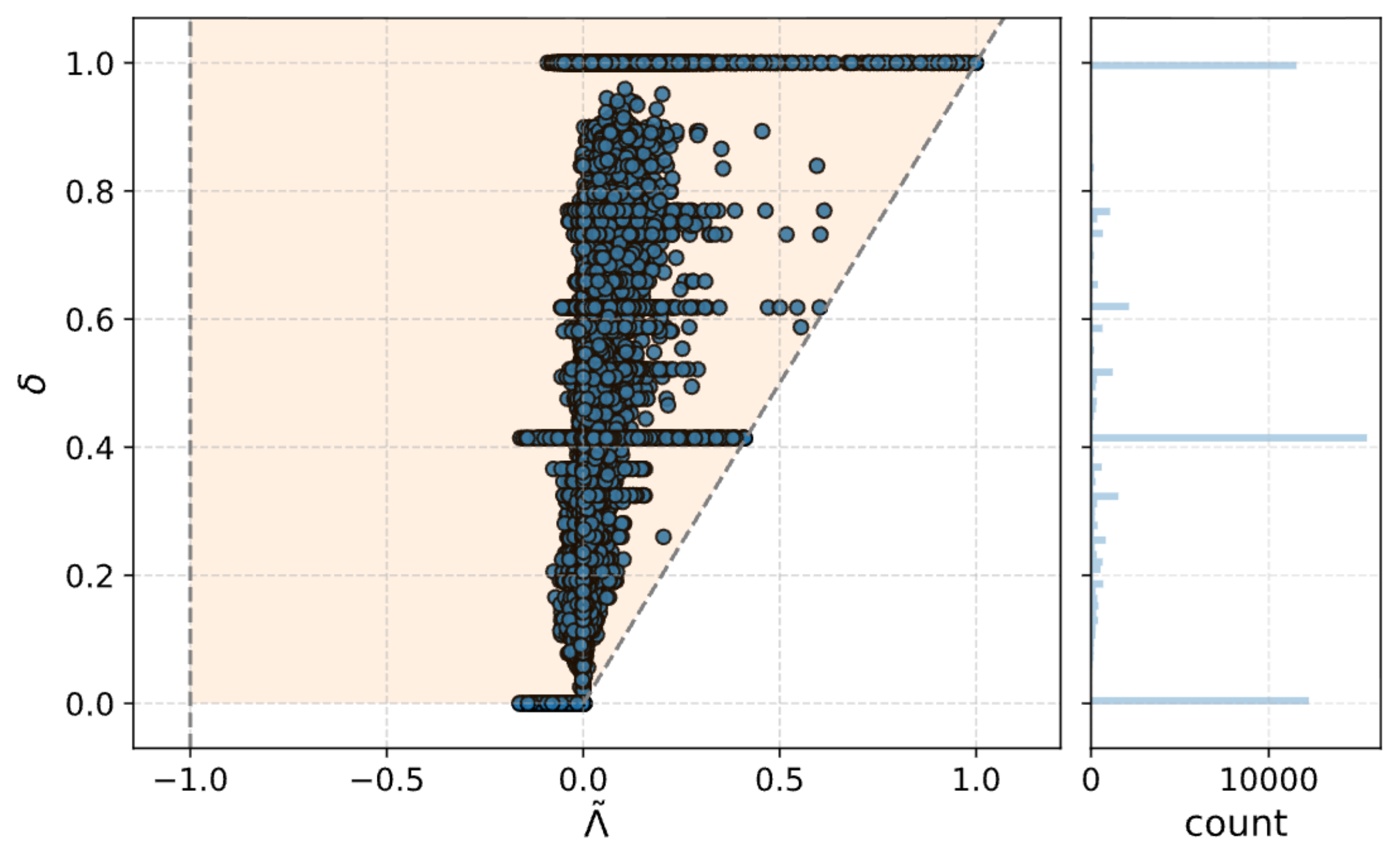}
    \end{subfigure}
    
    \caption{Growth parameter $\delta=\alpha-1$ versus normalized growth rate 
$\tilde{\Lambda}=\Lambda/\|\Sm\|_{\bm{\kappa}}$ for random autonomous, unambiguous CRNs with 
$2 \le |\mathcal{S}| \le 12$ and $2 \le |\mathcal{R}| \le 12$
and $M^+=M^-=2$. 
Left: autoinhibitory networks ($\alpha < 1$, $\delta < 0$). 
Right: autocatalytic networks ($\alpha > 1$, $\delta > 0$) and networks which admit a steady-state flux ($\alpha = 1$, $\delta = 0$). 
For $\alpha < 1$, all points lie within the expected strip $-1 \le \tilde{\Lambda} \le 0$, and for $\alpha \ge 1$, the rigorous bound $-1 \le \tilde{\Lambda} \le \delta$ holds throughout. 
Furthermore, most autoinhibitory networks satisfy $ \delta \leq \tilde{\Lambda} \leq 0$, while rare red outliers illustrate that the lower bound is not universal. The marginal distribution of $\delta$ is shown along the right side of each panel. The distribution exhibits accumulation near specific values, as discussed in the main text and Fig.~\ref{fig:core_labelled}.
}
    \label{fig:shrinkgrow}
\end{figure*}

\begin{figure*}[t]
    \centering
    
    \begin{subfigure}{0.47\textwidth}
        \centering
        \includegraphics[width=\linewidth]{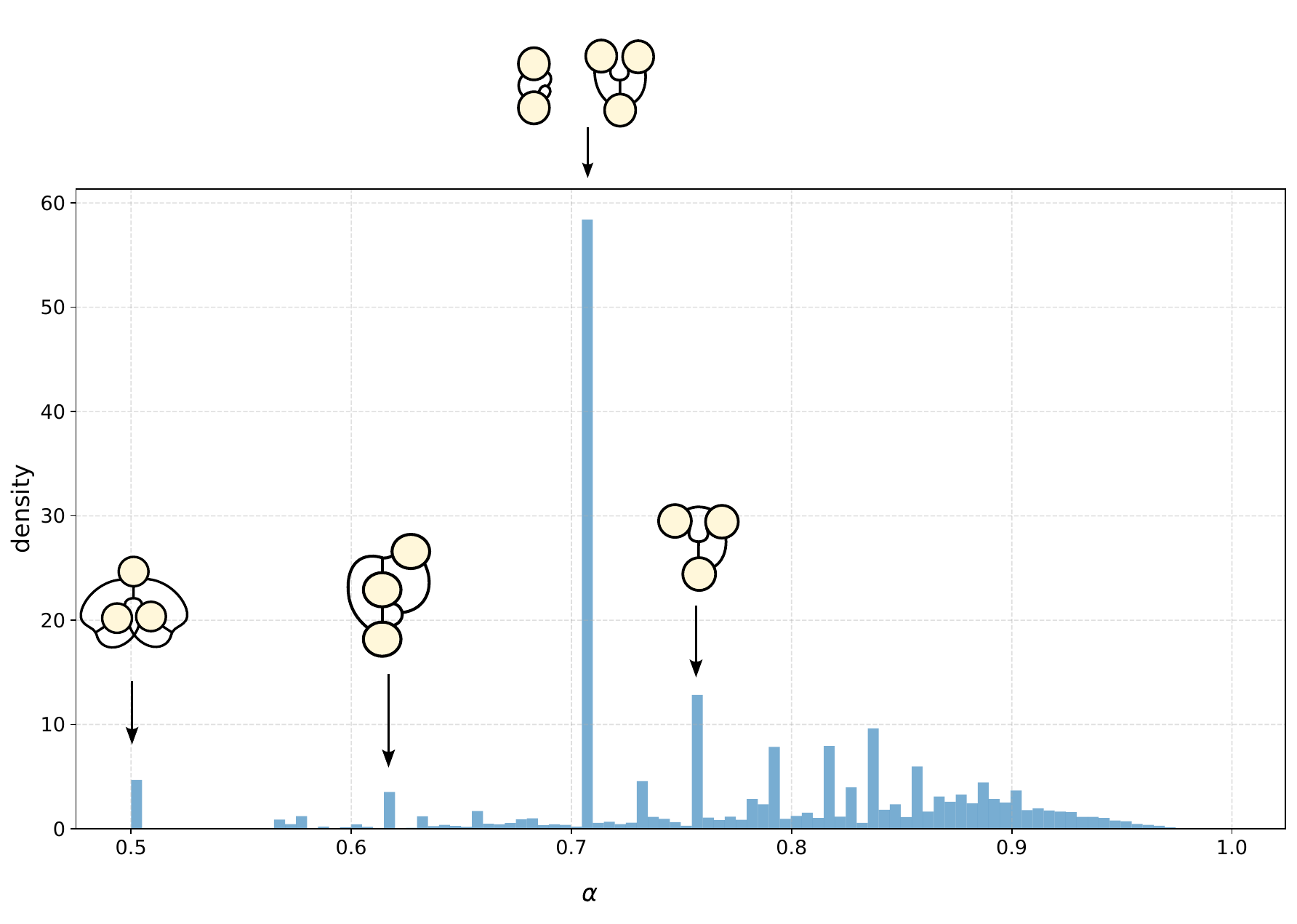}
    \end{subfigure}
    \hfill
    \begin{subfigure}{0.47\textwidth}
        \centering
        \includegraphics[width=\linewidth]{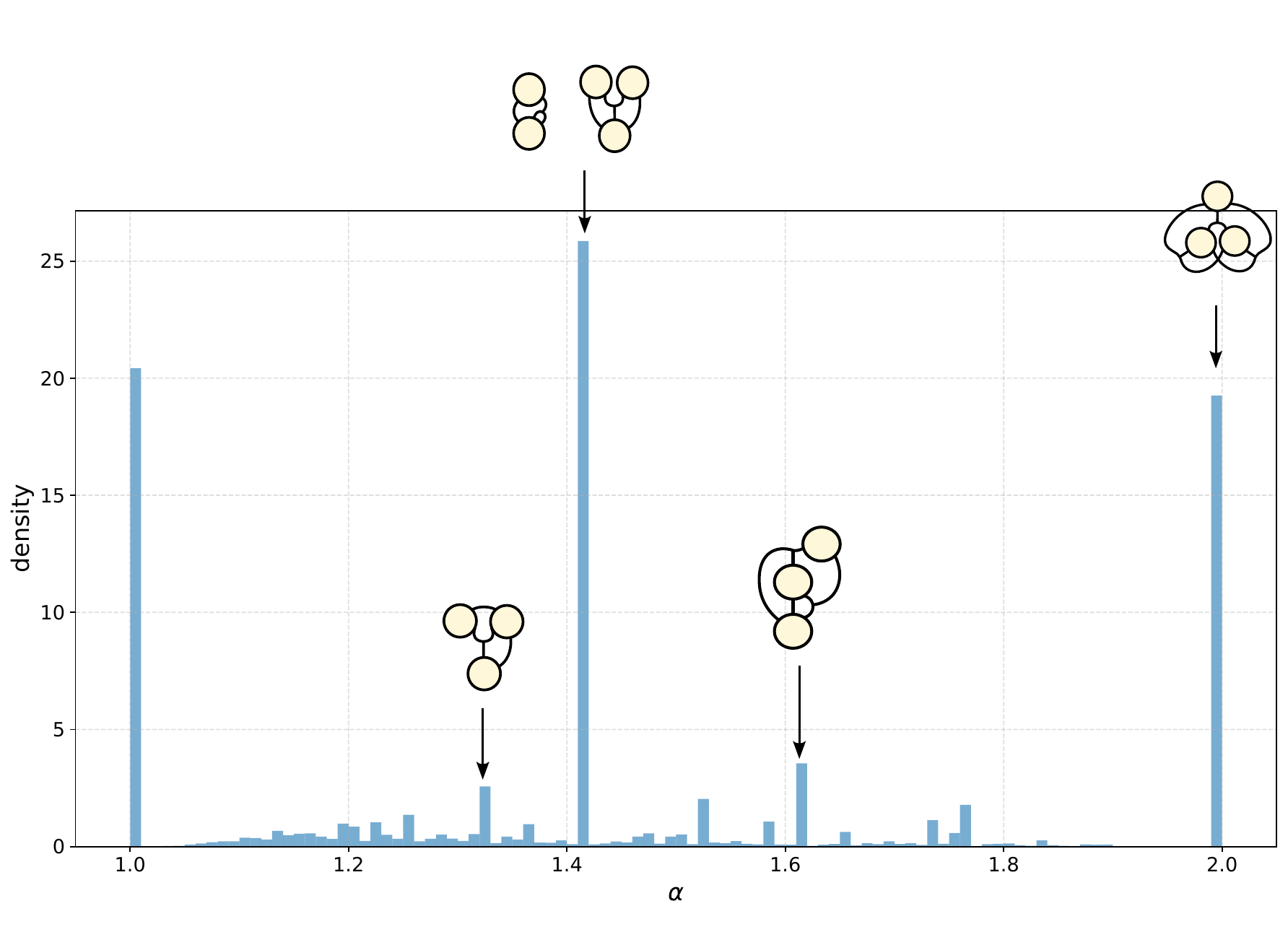}
    \end{subfigure}
    
    \caption{
Density of $\alpha$ over random networks from Fig.~\ref{fig:shrinkgrow} with motif-labelled peaks. 
Left: autoinhibitory networks with $\alpha < 1$. 
Right: autocatalytic networks with $\alpha > 1$ and networks exhibiting a steady state with $\alpha = 1$. 
Peaks are labeled by core types that share the same value of $\alpha$. 
For autocatalytic networks, some peaks correspond to autocatalytic cores,
while other peaks correspond to non-minimal structures, such as the peak at $\alpha=1$ for non-autocatalytic structures or the peak at $\alpha=2$.  
For autoinhibitory networks, some peaks correspond to the reciprocal autoinhibitory cores, obtained by reversing the reaction arrows from their autocatalytic definitions (see Table \ref{tab:MAFcores}).
}  \label{fig:core_labelled}
\end{figure*}


\textit{Topological bounds on $\Lambda$---}
Consider autonomous mass-action reaction networks that admit at least one fixed point $\q^*$ for which every species has non-zero abundance, i.e.~$q^*_s > 0$ for every $s$.
Since the vector of individual volumes \(\w\) is strictly positive, there exists
\(\epsilon>0\) such that \(\w \geq \epsilon\, \mbf{1}\), where \(\epsilon\) has
dimensions of volume. For each reaction \(r\), we define the rescaled rate
constant
\begin{align}
    \kappa_r := \frac{k_r}{\epsilon^{|r|-1}},
    \label{eq:rescaled_k}
\end{align}
which has dimensions of frequency. Here
\(|r|:=\sum_s (\Sm)_s^r\) denotes the order of reaction \(r\).

 For such networks, we obtain that
\(\Sm  \jj(\q^*) \leq \|\Sm\|_{\bm{\kappa}}  \phantom{\cdot}\q^* 
  \), where the kinetic--stoichiometric prefactor is
\begin{equation}
\label{eq:norm}
\|\Sm\|_{\bm{\kappa}} \;:=\; \max_{s\in\mathcal{S}}\;\sum_{r\in\mathcal{R}}(\Sm)^r_s\,\kappa_r.
\end{equation}
From this relation, we obtain a universal loose lower-bound \(
- \|\Sm\|_{\bm{\kappa}} < \Lambda.\)
We then show that if the MAF of a network is not greater than one, the network must either be steady or shrinking, i.e.\ $\alpha \leq 1 \implies \Lambda \leq 0$.
Finally, we show that growing networks satisfy
\(\;\Lambda \;\le\; \|\Sm\|_{\bm{\kappa}}\,(\alpha-1).\)
Collecting the above relations in a single equation, we get the main result of this work
\be
- \|\Sm\|_{\bm{\kappa}} \;<\; \Lambda \;\leq\; \max\left(0,\|\Sm\|_{\bm{\kappa}}\,(\alpha-1)\right). \label{eq:main}
\ee
The upper bound is saturated for autocatalytic networks with unimolecular inputs and equal rate constants. For these networks, furthermore, $(\alpha - 1)$ coincides with the largest eigenvalue of the stoichiometric matrix $\St$. Full details of these results are provided in the SM.

Equation~\eqref{eq:main} turns the MAF \(\alpha\) into a sharp, parameter-sparse descriptor of growth. 
First, it yields a strict \emph{decay certificate} for non-autocatalytic networks: if \(\alpha \leq 1\), then \(\Lambda \leq 0\) independently of the rate constants and of the choice of stable fixed point. 
Second, it imposes an explicit \emph{upper bound} for autocatalytic networks: even under arbitrary kinetic tuning, the balanced growth rate cannot exceed \(\|\Sm\|_{\bm{\kappa}}(\alpha-1)\). While Eq.~\eqref{eq:main} implies that any growing network must be autocatalytic, i.e.~\(\Lambda>0 \implies \alpha>1\), the converse does not hold: an autocatalytic network need not exhibit positive growth. In other words, the sign of \(\Lambda\) fundamentally depends on the kinetics when $\alpha>1$.

 Observe that the rescaled rate constants $\kappa_r$ correspond to choosing units of volume such that the smallest individual volume is unity, \(\epsilon=1\). When bounds on the rescaled rate constants are available, Eq.~\eqref{eq:main} leads to a clear separation between stoichiometric and kinetic structure of the network. Indeed, the volume growth rate $\Lambda$, rescaled by $K=\max_r \kappa_r$, is then bounded by quantities that depend only on the network topology:
\begin{equation}
-  \|\Sm\|_1 
\;<\;
\frac{\Lambda}{K}
\;\leq\;
\,\max\,\!\bigl(0,\|\Sm\|_1(\alpha-1)\bigr),
\label{eq:conservative_bound}
\end{equation}
where $\|\Sm\|_1=\max_s \sum_r (\Sm)_s^{\,r}$.

\textit{Example (cont.)---} 
For the type-I core, under mass-action with $\w = [1,1]^\top$ (so $q_A+q_B = 1$) and $k_1 = k_2 = 1$, the growth rate solution satisfies
\begin{align}
\Lambda &= \sqrt{2}-1, & q_A^* &= \frac{\sqrt{2}}{\sqrt{2}+1}, & q_B^* &= \frac{1}{\sqrt{2}+1}.   
\end{align}
For this assignment $\|\Sm\|_{\bm{\kappa}} = \max(1,1) = 1$, and from Equation \ref{eq:main}, we have the upper bound
\( \Lambda \leq 1\cdot(\sqrt{2}-1),\)
which is exactly saturated. Also, observe that $\alpha = \sqrt{2}$ is simply the highest eigenvalue of $\Sp$, and $(\sqrt{2}-1)$ is the highest eigenvalue of $\St$.

\textit{Numerical tests on random autonomous networks---} 
We tested the bounds on randomly generated autonomous, unambiguous CRNs with $2 \le |\mathcal{S}| \le 12$ and $2 \le |\mathcal{R}| \le 12$ (SM).
We also vary the maximum reaction order for products, $M^+$, and reactants, $M^-$, where
\( M^\pm := \max_{r } \sum_{s }(\St^\pm)_s^r.\)
For each instance, we computed \(\alpha\) from \eqref{eq:MAF} using computational implementations in \cite{sargent2019neumann,blanco2025identifyingselfamplifyinghypergraphstructures}. When necessary, we also used integration of the rate equation \eqref{eq:rateeq_main} to find a stable fixed point \(\q^*\) following \cite{lin2020origin}, and extracted \(\Lambda\) from \eqref{eq:LambdaDef}. Using $\delta \equiv \alpha - 1$ and $\tilde{\Lambda} \equiv \Lambda/\|\Sm\|_{\bm{\kappa}}$, Equation \ref{eq:main} can be recast as
\be
-1 < \tilde{\Lambda} \leq \max(0,\delta).
\ee
Figure~\ref{fig:shrinkgrow} confirms that all networks satisfy this bound. The majority of autocatalytic networks with (\(\alpha>1\)) are far from the limit $\delta \simeq 0$, and  their growth rate stays far from the lower bound $\tilde{\Lambda}=-1$ and is  positive. Thus, these networks have a built-in tendency for growth.
The majority of autoinhibitory networks (\(\alpha<1\)), satisfy the empirical envelope
\(
 \delta \leq \tilde{\Lambda}.
\)
This lower bound is much tighter than the one of Eq. \eqref{eq:main}, but it is \emph{not universal}:  it can fail. 
In that case, one obtains the outliers shown in red in the right figure of Fig.~\ref{fig:shrinkgrow}. As explained in SM, these outliers mainly appear in the region near $\delta \simeq 0^-$. 

A striking feature of the left column of Fig.~\ref{fig:shrinkgrow} for maximum reaction orders $M^+=M^-=2$ is that the distribution  of $\delta = \alpha-1$ values concentrates on a discrete set of values. 
In Fig.~\ref{fig:core_labelled}, we have labeled the peaks in $\alpha$ together with a representation of the cores with the same $\alpha$ among the five cores of reversible autocatalytic networks. We observe that a large number of networks concentrate on the MAF value of type I or type III cores, which is $\sqrt{2}$ or on the value 2. A large degeneracy of networks is observed among networks with $\alpha \ge 1$, at values $\alpha=1$, and at $\alpha=2$, which coincides with the value of $M^+$. In SM, we provide more details about the sampling procedure, and we show that the MAF $\alpha$ satisfies $1/M^- \leq \alpha \leq M^+$. For higher reaction orders $M^+$ and $M^-$, the peaks flatten out and the distribution approaches a continuous spectrum (SM). 

\textit{Discussion and outlook---}
Our main result, Eq.~\eqref{eq:main}, bounds the balanced-growth rate by a topological factor times a kinetic--stoichiometric prefactor. The topological factor is the MAF \(\alpha\), a purely structural quantity fixed by network topology, whereas the kinetic--stoichiometric prefactor, \(\|\Sm\|_{\bm{\kappa}}\), depends on the chosen kinetics. 
Operationally, one can compute \(\alpha\) from stoichiometry~\cite{blanco2025identifyingselfamplifyinghypergraphstructures,sargent2019neumann,de_martino_von_2012} and estimate \(\|\Sm\|_{\bm{\kappa}}\) from input stoichiometries and rough kinetic scales, thereby obtaining an upper bound on the achievable growth rate. 

For a fixed topology, changing rate constants varies the realized growth rate
\(\Lambda\) continuously, but only within limits set by the MAF \(\alpha\).
Thus kinetic tuning can optimize performance within a fixed admissible
window, whereas structural changes---such as adding or removing reactions,
merging motifs, or rewiring autocatalytic components---can change
\(\alpha\) itself and shift the bound. This separation suggests a minimal
picture for chemical evolution \cite{qian_chemical_2010,sole_fundamental_2024,jain_model_2001}: neutral or adaptive kinetic changes explore
growth rates at fixed MAF, while rarer topological changes alter the
network-level growth potential. 

More practically, this viewpoint turns \(\alpha\) into a target for network
engineering. A key next step is to develop composition rules for the MAF:
which elementary operations, such as adding a reaction, merging species, or
coupling autocatalytic motifs, increase or decrease \(\alpha\), and by how
much? Such rules would connect topological capability, encoded by \(\alpha\),
to dynamical realization, encoded by the steady flux, and identify the
network motifs that control growth performance
\cite{sarkar_design_2019,despons_stability_2025}.


Our result also has practical implications for open chemical reactors \cite{marehalli_srinivas_thermodynamics_2024}. In a continuously stirred reactor with dilution rate \(d\), after possible transients the population scale obeys $\dot{\Omega} = (\Lambda-d)\,\Omega$. Sustained growth therefore requires \(\Lambda>d\); equivalently, dilution rates above \(\Lambda\) wash out the system. Combined with Eq.~\eqref{eq:main}, this gives a necessary structural condition for persistence against dilution: even with optimal kinetic tuning, survival at dilution \(d\) is excluded unless \(\alpha>1\) and \(d\) lies below the corresponding growth bound \cite{kolchinsky2025thermodynamics}.

Finally, balanced growth is an asymptotic statement. In finite populations, demographic noise and rare-event switching can induce extinction or transitions between composition attractors even when a stable \(\q^*\) exists deterministically \cite{singh2023multistable, lazarescu_large_2019}. Extending the framework to stochastic kinetics would quantify the reliability of topologically permitted growth bounds and yield risk-aware structural criteria, such as expected persistence times under fluctuations.

\textit{Acknowledgments---}
P.G.\ acknowledges Eric Smith, Nathaniel Virgo, and Atsushi Kamimura for discussions. J.W.\ acknowledges Chunqiu Xia for computational resources. D.L.\  acknowledges Barnabé Ledoux and Paul Perret for discussions and help with numerical computations.
 P.G.\ was funded by JST CREST (JPMJCR2011) and JSPS Grant No.\ 25H01365.
This research was partially conducted while visiting the Okinawa Institute of Science and Technology (OIST) through the Theoretical Sciences Visiting Program (TSVP) Thematic Program on Biological Information Processing (TP24BI).
\putbib
\end{bibunit}


\clearpage
\onecolumngrid

\title{Supplemental Material for\\
``Topological bounds on the dynamical growth rate of chemical reaction networks''}

\maketitle

\setcounter{equation}{0}
\setcounter{figure}{0}
\setcounter{table}{0}
\renewcommand{\theequation}{S\arabic{equation}}
\renewcommand{\thefigure}{S\arabic{figure}}
\renewcommand{\thetable}{S\arabic{table}}
\renewcommand{\thesection}{S\arabic{section}}
\begin{bibunit}

\section*{ Supplemental Material for ``Topological bounds on the dynamical growth rate of chemical reaction networks''}

\section*{ A. Structural setting and definitions}
\subsection*{ A1. Notation}
Vectors are represented in boldface. The $i$th component of a vector $\w$ is denoted with $w_i$. A vector $\w$ is said to be greater than zero, denoted as $\w > \0$, if $w_i \geq 0$ for all $i$ but $\w \neq \0$. A vector $\w$ is said to be strictly greater than zero, denoted as $\w \gg \0$, if $w_i > 0$ for all $i$. Each vector is assumed to be a column vector. Matrices are denoted in blackboard bold. The entry of the $i$th column and $j$th row of the matrix $\mbb{A}$ is denoted as $\mbb{A}^i_j$. A matrix $\mbb{A}$ is said to be nonnegative if all its entries are nonnegative. 

\subsection*{ A2. Chemical reaction network}
A chemical reaction network (CRN) with species set $\mathcal{S}=\{1,\dots,|\mcl{S}|\}$ and reaction set $\mathcal{R}=\{1,\dots,|\mcl{R}|\}$ is encoded by two nonnegative integer matrices $|\mcl{S}| \times |\mcl{R}|$ matrices
\begin{equation}
(\Sm)_{s}^r\ge 0,\qquad (\Sp)_{s}^r\ge 0,
\end{equation}
collecting the input (reactant) and output (product) stoichiometric coefficients of reaction $r$ on species $s$. In this work, we restrict the stoichiometries to be integers. 
The stoichiometric matrix is
\begin{equation}
\St = \Sp-\Sm.
\end{equation}
For any nonnegative flux vector $\x \geq \0 \in\mbb{R}^{|\mcl{R}|}_{\ge 0}$, the species-wise input and output fluxes are
\begin{equation}
\Sm\x \geq \0 \in\mbb{R}^{|\mcl{S}|}_{\ge 0},\qquad \Sp\x \geq \0 \in\mbb{R}^{|\mcl{S}|}_{\ge 0}.
\end{equation}

\subsection*{ A3. Autonomous network}
A CRN $(\Sp,\Sm)$ is \textbf{autonomous} if \cite{gagrani2024polyhedral,blanco2025identifyingselfamplifyinghypergraphstructures}:
\begin{enumerate}[label=(A\arabic*)]
    \item \label{eq:spc_consume} All species are consumed by some reaction.
    \[
    \forall s \in \mcl{S}, \exists r \in \mcl{R} : (\Sm)^r_s >0. 
    \]
    \item \label{eq:spc_produce} All species are produced by some reaction.
    \[ 
    \forall s \in \mcl{S}, \exists r \in \mcl{R} : (\Sp)^r_s >0. \]
    \item \label{eq:rxns_consume} All reactions consume some species.
    \[
    \forall r \in \mcl{R}, \exists s \in \mcl{S} : (\Sm)^r_s >0.  
    \]
    \item \label{eq:rxns_produce} All reactions produce some species.
    \[
    \forall r \in \mcl{R}, \exists s \in \mcl{S} : (\Sp)^r_s >0. 
    \]
\end{enumerate}
A network where each reaction consumes and produces each species, i.e.~which satisfies conditions \ref{eq:spc_produce} and \ref{eq:rxns_consume}, will be called \textit{semi-autonomous}.

\subsection*{ A4. Autocatalytic network}
A CRN $(\Sp,\Sm)$ is \textbf{autocatalytic} if:
\begin{enumerate}[label=(AC\arabic*)]
    \item \label{prop:autonomy} \textit{Autonomy:} The CRN is autonomous (satisfies \ref{eq:spc_consume}-\ref{eq:rxns_produce}).
    \item \label{prop:productivity} \textit{Productivity:} There exists a flux vector $\x$ such that all species are net-produced positively, 
    \be
     \exists \x > \0: \;  \St \x \gg \0 \iff \Sp \x \gg \Sm \x \iff \alpha(\x) > 1. 
    \ee
\end{enumerate}
 Note that requiring all reactions to be \textit{unambiguous}-i.e.\ no species are shared between the reactants and products of any reaction-recovers the corresponding definitions of autonomy and autocatalysis in \cite{blokhuis2020universal}. The results in this work does not require this assumption.
 
 \subsection*{ A5. Reducible and irreducible network}
A CRN is \textbf{reducible} \cite{gale1989theory,sargent2019neumann} if there exists a semi-autonomous subnetwork inside the whole network, that is able to produce every species in itself without consuming species from outside. Mathematically, a CRN $(\mcl{S},\mcl{R})$ is reducible if there exists $(\mcl{S}',\mcl{R}')$ such that
\begin{align}
     \forall s \in \mcl{S}', \exists r \in \mcl{R}' : &
     (\Sp)^r_s >0,\\
     \forall r \in \mcl{R}', \forall s \in \mcl{S}/\mcl{S}' : &(\Sm)^r_s =0.    
\end{align}
A CRN is said to be \textbf{irreducible} if it is not reducible.

\section*{ B. Balanced-growth regime}

\subsection*{B1. Composition dynamics }
For species population $\n$, relative weight vector $\w>\0$, and scale $\Omega= \w^T \n$, we define the relative population vector 
\begin{align}
    \q &= \frac{\n}{\Omega}= \frac{\n}{\w^T \n}. \label{eq:concentration}
\end{align}
Consider CRN dynamics
\begin{align}
    \dot{\n}(t) &= \St \, \J(\n(t)), \label{eq:dxdt}
\end{align}
that satisfies 
\begin{align}
    \frac{\J(\n)}{\Omega} = \frac{\J(\q \Omega)}{\Omega} &= \jj(\q), \label{eq:scaling}
\end{align}
Taking a time derivative of $\Omega$ and using the change of coordinates readily yields
\be    \dot{\Omega}
    = \Omega\, (\w^T \St \, \jj(\q)).\ee
Substituting $\n = \q \, \Omega$ and using the chain-rule in Equation \ref{eq:dxdt} yields the composition equation
\begin{align}
    \dot{\q} &= \St \, \jj(\q) - (\w^T \St \, \jj(\q)) \, \q. \label{eq:dqdt} 
\end{align}
The fixed point of the above equation yields the relation stipulated in Equation \ref{eq:LambdaDef} of the main text.

\subsection*{B2. Recovering the dilute constant-volume approximation }
Consider a CRN with species set ${X_1,\ldots,X_N}$ where the dynamics are given by the linear-volume mass-action form in Eq.~\ref{eq:mass-action_prop}
\begin{align}
\J(\n) 
    &=
    k_r \Omega \prod_{s \in \mcl{S}} \left( \frac{n_s}{\Omega} \right) ^{(\Sm)^r_s}.
\end{align}
Suppose $X_1$ is a solvent species that is neither consumed nor produced in each one-way reaction, then every reaction $r$ is of the form:
\begin{align}
    r: \; \sum_{s=2}^{N} (\Sm)_s^r X_s \to
    \sum_{s=2}^{N} (\Sp)_s^r X_s.
\end{align}
The volume of the system is given by 
\be
\Omega = \sum_{s=1}^N w_s n_s.
\ee
Assume the dilute limit, where the abundance of the solvent is much greater than other species, $n_1 \gg n_s $ for $s\geq 2$. Thus,
\be
\Omega \approx w_1 n_1.
\ee
Furthermore, since the solvent is not net-produced by any reaction, $d\log(\Omega)/dt\approx 0$, and the growth rate is approximately zero. This yields, the standard constant-volume mass-action form where $\Omega$ is a constant proportional to the abundance of the solvent.

\section*{ C. Maximum Amplification Factor (MAF) results}
\subsection{ C1. Reformulation of MAF }
The MAF can be reformulated as the solution to the following optimization problem:
\bea
\alpha(\Sp,\Sm) = & \hspace{-5em}
\max\limits_{\x > \0 } \qquad \alpha(\x) \nn\\
& \text{such that } \quad \Sp \x  \geq \alpha(\x) \phantom{\cdot}\Sm \x. \label{eq:reformulation_MAF}
\eea
 For solver implementations of the above, see \cite{sargent2019neumann,blanco2025identifyingselfamplifyinghypergraphstructures}.

\subsection*{ C2. MAF bounds for autonomous networks}
Von Neumann in \cite{neumann1945model} and
Theorem 9.8 in \cite{gale1989theory} rigorously prove that $0 < \alpha < \infty$ for semi-autonomous networks. Since autonomous networks are also semi-autonomous, the bounds trivially extend to them as well. While detailed proofs of the statement can be found there, we sketch a heuristic derivation of the same. 

Recall from Equation \ref{eq:MAF} that
\begin{equation}
\alpha \;=\; \max_{\x>\bm{0}}\;\min_{s\in\mathcal{S}}
\frac{(\Sp\x)_s}{(\Sm\x)_s}.
\end{equation}
Since, by property \ref{eq:spc_produce}, every species is produced by at least one reaction, there exists a feasible flux $\x > \0$ such that the numerator $\Sp \x \gg \0$. Thus, for such a flux, \((\Sp \x)_s/(\Sm \x)_s>0\) for every $s$ and, in particular, the minimum over all species is greater than zero. Furthermore, by property \ref{eq:rxns_consume}, every reaction consumes and thus, for the same $\x$, $\Sm \x > \0$ (as not all elements $(\Sm \x)_s$ can be simultaneously zero). Thus, there must exist at least one species $s$ for which \((\Sp \x)_s/(\Sm \x)_s< \infty\) which implies that minimum over all species must also be less than infinity. Since $\alpha$ is the maximum over such values, it must also satisfy $0<\alpha<\infty$. 

\subsection*{ C3. MAF of reducible networks}

Von Neumann in \cite{neumann1945model} and
Theorem 9.10 in \cite{gale1989theory} prove that there exists a unique MAF for irreducible semi-autonomous networks. Since autonomous networks are also semi-autonomous, irreducible autonomous networks have a unique MAF. 

We now turn to the case of reducible networks. If the network decomposes into disjoint irreducible components that do not exchange mass—equivalently, if $(\Sp,\Sm)$ can be brought into block-triangular form after a suitable permutation—then the maximum amplification factor is simply the maximum of the MAFs of the individual irreducible blocks.

More generally, a reducible network may consist of an irreducible subnetwork whose species feed one or more downstream subnetworks. In this case, any flux through downstream reactions necessarily increases the consumption of species in the core without providing compensating production within it. As a result, activating downstream subnetworks cannot increase the uniform production-to-consumption ratio across species, and can only reduce the achievable amplification factor. Optimal fluxes therefore concentrate on the dominant irreducible component.

In this sense, $\alpha$ functions as a “best-module” structural score and is determined solely by the irreducible component with the highest MAF.

\subsection*{ C4. MAF of minimal networks}
\label{sec:minimal_nets}
Observe from the second formulation of the MAF (Equation \ref{eq:reformulation_MAF}) that the optimal solution $\x>\0$ for a CRN $(\Sp,\Sm)$ satisfies
\be \Sp \x \;\geq\; \alpha\, \Sm \x. \ee
 While, in general, the above equation only holds as an inequality, the equality must be satisfied for some  species and reactions otherwise $\alpha$ would not be the maximum value that satisfies the inequality. Given the subnetwork $(\rSp,\rSm)$ and $\rx$, we obtain the following \textit{generalized eigenvalue problem}:
\begin{align}
    \rSp \rx = \alpha \, \rSm \rx, \label{eq:GEP}
\end{align}
with the constraint $\rx > \0$. 
The relation between solutions of Eq.~\eqref{eq:reformulation_MAF} and the generalized eigenvalue problem is studied in Ref.~\cite{thompson1971neumann}.

An autonomous network is \textit{minimal} if there is no subnetwork that is also autonomous. Since there is no species that is never consumed, such a network must also be irreducible. It has been shown that minimal autonomous networks have an equal number of species and reactions \cite{gagrani2024polyhedral,blokhuis2020universal,vassena2024unstable}, and that their input matrix can be rearranged into a diagonal form. Clearly, then, the network is irreducible and $\Sm$ is non-singular. Using the results in \cite{thompson1971neumann}, the problem of finding MAF reduces to the generalized eigenvalue problem of Eq.~\eqref{eq:GEP}, which further reduces to the standard eigenvalue problem
\be (\Sm)^{-1}\Sp  \x = \alpha  \, \x,\ee
where $\alpha$ is the highest eigenvalue of $(\Sm)^{-1}\Sp$ with a corresponding nonnegative Perron-Frobenius eigenvector $\x$.

\section*{ D. Bounds on the dynamic growth rate}

\subsection*{ D1. Assumptions}
We assume an autonomous network under mass--action kinetics Eq.~\eqref{eq:mass-action_prop} with positive rate constants and the existence of at least one fixed point of Equation \ref{eq:dqdt} in the positive orthant $\q^{*}\gg \0$, ensuring that the growth is exponential with a rate defined by Eq.~\eqref{eq:LambdaDef}. 

The vector of individual volumes must be positive, thus $\exists \epsilon>0, \, s.t. \, \w \ge \mbf{1} \, \epsilon$.
Since $\w^{T}\q = 1$, using $\max_s q_s \leq \sum_s q_s$, it follows that 
\begin{equation}
    1 \geq \epsilon \mbf{1} \cdot \q \geq \epsilon \max_s q_s,
\end{equation}
which yields that 
$\max_s q_s \leq 1/\epsilon$. 

\subsection{ D2. Kinetic-stoichiometric prefactor}

From the form of the propensity of Eq.~\eqref{eq:mass-action_prop}, 
\be
    (\Sm  \jj(\q))_s := 
       \sum_{r \in \mcl{R}} (\Sm)^r_s k_r  \prod_{t \in \mcl{S}} q_t^{(\Sm)^r_t}.
\ee
Given assumption \ref{eq:spc_consume}, for each species $s$, there exists a reaction $r$ such that $(\Sm)^r_s \geq 1$.
For such a pair of $s$ and $r$, we rewrite $\q^{(\Sm)^r_s} = q_s p_r^s(\q)$, where $p_r^s(\q)$ is a monomial of the concentration coordinates. This yields,
\begin{align}
     (\Sm  \jj(\q))_s
     &= 
     q_s \sum_{r: (\Sm)_s^r>0} (\Sm)^r_s k_r p^s_r(\q).
\end{align}
Let us denote the order of reaction $r$ by $|r|$ where $|r|:= \sum_{s} (\Sm)^r_s$. Observe that $p^s_r$ is a monomial of order $|r|-1$.
Since all the concentrations are upper bounded by $1/\epsilon$, for every $r$, $p^s_r(\q) \leq 1/\epsilon^{|r|-1}$. Putting these together, we get
\begin{align}
    (\Sm  \jj(\q))_s \leq q_s \sum_{r} (\Sm)^r_s \frac{k_r}{\epsilon^{|r|-1}} \leq q_s \max_{t \in \mcl{S} } \left( \sum_{r } (\Sm)^r_t \frac{k_r}{\epsilon^{|r|-1}}\right).
\end{align}
Using 
\begin{equation}
\|\Sm\|_{\bm{\kappa}} \;:=\; \max_{s\in\mathcal{S}}\;\sum_{r\in\mathcal{R}}(\Sm)^r_s\,\frac{k_r}{\epsilon^{|r|-1}},
\end{equation}
we obtain the inequality
\be       
\label{eq:inequality_on_q}
    \Sm  \jj(\q) \leq \|\Sm\|_{\bm{\kappa}}  \phantom{\cdot}\q. 
\ee

\subsection{ D3. Universal lower bound}
By Equation \ref{eq:LambdaDef}, \(\St \, \jj(\q^*) = \Lambda\, \q^*\), for every species $s$
\begin{align}
    \Lambda = \frac{(\St \, \jj(\q^*))_s}{q^*_s} 
    = \frac{(\Sp \jj(\q^*))_s - (\Sm \jj(\q^*))_s }{q^*_s}. \label{eq:lambda_SM}
\end{align}
Since $\q^*$ is in the positive-orthant, $\jj(\q^*) \gg \0$. Furthermore, by assumptions \ref{eq:spc_produce} and \ref{eq:rxns_produce}, every reaction produces and every species is produced by some reaction, yielding $\Sp \jj(\q^*) \gg \0$.
This means that
\begin{align}
    \Lambda > -\frac{ (\Sm \jj(\q^*))_s }{q^*_s}   \geq - \|\Sm\|_{\bm{\kappa}},
\end{align}
where Eq.~\eqref{eq:inequality_on_q} is used in the last line. 

\subsection*{ D4. $\alpha \leq 1$ implies $\Lambda \leq 0$ }
From the definitions of MAF, Equations \ref{eq:AF} and \ref{eq:MAF} in the main text, and Equation \ref{eq:reformulation_MAF} in SM, we obtain the following implication and its contrapositive:
\begin{align}
    \exists \x > \0:&\; \St \x \gg \0 \implies 1 < \alpha, \nonumber\\
    \alpha \leq 1 & \implies
    \nexists \x > \0 :\; \St \x \gg \0. \label{eq:ineq_MAF}
\end{align}
In words, \(\alpha \leq 1\) implies that there is no flux \(\x > \mathbf{0}\) such that \(\St \x \gg \mathbf{0}\). 
This is equivalent to $\forall \x, \exists s \;  \rm{ s. t. } \; (\St \x)_s \le 0$. Now, we can use $\x=\jj(\q^*)$ in Eq. \ref{eq:lambda_SM} together with the corresponding $s$ in that equation. Then, since $q_s \ge 0$, it follows that $\Lambda \le 0$. In other words, the network must either be steady or shrinking.



\subsection*{ D5. Upper bound on growing networks}
For growing networks, for which $\Lambda>0$,
\begin{align}
 \Lambda 
    &= \min_{s \in \mcl{S}} \frac{(\St \, \jj(\q^*))_s}{(\Sm \jj(\q^*))_s} \frac{(\Sm \jj(\q^*))_s}{q^*_s} \nonumber\\
    & \leq 
 \|\Sm\|_{\bm{\kappa}}   \min_{s \in \mcl{S}} 
    \frac{ \left( \St \, \jj(\q^*) \right)_s}{\left( \Sm \jj(\q^*) \right)_s}, \nonumber\\
    & \leq 
    \|\Sm\|_{\bm{\kappa}}
    \max_{\x > \0 } \min_{s \in \mcl{S}} \frac{(\St \x)_s}{(\Sm \x)_s} \nonumber\\
    & \leq 
    \|\Sm\|_{\bm{\kappa}} \left(
    \max_{\x > \0 } \min_{s \in \mcl{S}} \frac{(\Sp \x)_s}{(\Sm \x)_s} -1\right) \nonumber\\
    & \leq   \|\Sm\|_{\bm{\kappa}} \left( \alpha(\Sp,\Sm)-1 \right),
\end{align}
where in the second line $\Lambda>0$ and Equation \ref{eq:inequality_on_q} are used, and the definition of MAF (Eq.~\ref{eq:MAF}) is used to obtain the final line. Observe that, by assumptions \ref{eq:spc_consume} and \ref{eq:rxns_consume}, every reaction consumes and every species is consumed by some reaction, yielding $\Sm \jj(\q^*) \gg \0$ and a positive denominator for every $s$.

\subsection{ D6. Bounds on unimolecular networks}
From the discussion in Section C3, for minimal autonomous networks with unimolecular inputs, the input matrix is the identity, $\Sm = \mbb{I}$, and the generalized eigenvalue problem reduces to
\(
\Sp \x = \alpha \x,
\)
where $\alpha$ is the Perron--Frobenius eigenvalue of $\Sp$. Furthermore, the same eigenvector $\x$ satisfies
\be
\St \x = (\Sp - \mbb{I}) \x = (\alpha - 1)\x,
\ee
from which we see that $(\alpha - 1)$ is the largest eigenvalue of the stoichiometric matrix $\St$.

Assume that all species contribute equally to the volume and $\epsilon=1$, thus $\w=\mbf{1}$. From Equation~\ref{eq:norm}, we get
\be
\|\Sm\|_{\bm{\kappa}} = \max_r k_r,
\ee
so the kinetic--stoichiometric prefactor reduces to the maximum rate constant. Thus, in the case where all rate constants are equal, $k_r = k \ \forall r \in \mcl{R}$ the inequality becomes an equality,
\be
\Lambda = k(\alpha - 1),
\ee
and the fixed point $\q^*$ is given by the Perron--Frobenius eigenvector $\x$ of $\St$, normalized such that $\w^T \x = 1$.

\subsection*{ D7. Non-universal lower bound}

\label{app:lower bound}

For shrinking networks, for which \(\Lambda < 0\), then:
\be
\label{eq:Shrink-netw1}
(\mathbb{S}^+ \jmath(\mathbf{q}^*))_s < (\mathbb{S}^- \jmath(\mathbf{q}^*))_s.
\ee

We also have that \(\Lambda\) can be expressed as:
\be
\Lambda = \min_{s \in \mathcal{S}} \frac{(\mathbb{S} \jmath(\mathbf{q}^*))_s}{q_s^*} = \min_{s \in \mathcal{S}} \left( \frac{(\mathbb{S}^+ \jmath(\mathbf{q}^*))_s}{q_s^*} - \frac{(\mathbb{S}^- \jmath(\mathbf{q}^*))_s}{q_s^*} \right).
\ee

Using the definition of the MAF $\alpha$, for the flux vector \(\jmath(\mathbf{q}^*)\), we have:
\be
\label{eq:Shrink-netw2}
\alpha \geq \min_{s \in \mathcal{S}} \frac{(\mathbb{S}^+ \jmath(\mathbf{q}^*))_s}{(\mathbb{S}^- \jmath(\mathbf{q}^*))_s} = \alpha(\jmath(\mathbf{q}^*)).
\ee

Defining a positive quantity
\be
\gamma := \alpha - \alpha(\jmath(\mathbf{q}^*)),
\ee
for each species \(s\), we can write:
\be
\label{eq:Shrink-netw3}
\frac{(\mathbb{S}^+ \jmath(\mathbf{q}^*))_s}{(\mathbb{S}^- \jmath(\mathbf{q}^*))_s} \geq \alpha - \gamma.
\ee
Note that \(\gamma\) tends to zero as \(\jmath(\mathbf{q}^*)\) as approaches the optimal flux vector.

Using the above inequality, we can write:
\be
\label{eq:Shrink-netw4}
\Lambda  \geq \min_{s \in \mathcal{S}} \left( (\alpha - \gamma - 1) \frac{(\mathbb{S}^- \jmath(\mathbf{q}^*))_s}{q_s^*} \right).
\ee

Now using the inequality of Eq.~\eqref{eq:inequality_on_q}, namely  $(\mathbb{S}^- \jmath(\mathbf{q}^*))_s / q_s^* \leq \|\mathbb{S}^-\|_k$, we have:
\be
\Lambda \geq (\alpha - \gamma - 1) \|\mathbb{S}^-\|_k.
\ee

For a majority of random networks, we empirically find
\be
\Lambda \geq (\alpha - 1) \|\mathbb{S}^-\|_k.
\ee
This holds when $\jmath(\mathbf{q}^*)$ is sufficiently close to the optimal flux vector,
so that $\gamma$ is negligible. However, there are some networks, especially those for which $\alpha-1$ is small, for which the above bound is broken. In other words, this lower bound is not universal. The networks that break the bound correspond to the outliers visible in Fig.~\ref{fig:shrinkgrow}.

\section*{ E. Numerical tests}

\subsection{ E1. Construction of the random networks}

We construct a random stoichiometric matrix $\St \in \mbb{Z}^{S\times R}$, where $S$ and $R$ respectively denote the number of species and reactions in the network. The integers $S$ and $R$ are randomly drawn in the set $[2,12]$, where $12$ is chosen arbitrarily to restrict the maximum system size.

Recall that the stoichiometric matrix is decomposed in the output and input nonnegative integer-valued matrices  $\St = \Sp - \Sm$. We define the product and reactant \textit{reaction order}, respectively, as
\be
\label{def_m+_m-}
 m_r^+ := \sum_{s \in \mcl{S}} (\Sp)^r_s, \quad m_r^- := \sum_{s\in \mcl{S}} (\Sm)^r_s. 
\ee
Let us also introduce a couple of integer parameters $(M^+, M^-)$ referred to as the maximum forward (resp. backward) reaction order such that
    \be
    m_r^+ \le M^+, \qquad m_r^- \le M^-
    \quad \text{for all } r\in \mcl{R}.
    \ee

The output and input matrices are sampled by the same procedure. Here, we describe it for the output matrix \(\Sp\). First, choose one reaction \(r^*\in[1,R]\) uniformly at random and set its reaction order to \(M^+\). For each remaining reaction \(r\), sample the reaction order \(m_r^+\) uniformly from \([1,M^+]\). Conditional on \(m_r^+\), construct the column \((\Sp)^r\) by repeatedly sampling species uniformly at random until the prescribed reaction order is reached. 
This yields the output matrix \(\Sp\). The input matrix \(\Sm\) is sampled analogously.

Finally, only networks satisfying the \textit{unambiguous} and \textit{autonomous} conditions are retained. A network is unambiguous if no reaction shares a species between its reactant and product complexes. Although the theoretical results of this work do not require unambiguity, ambiguous reactions are autonomous in the repeated species. Such reactions therefore contain an irreducible autonomous subnetwork supported entirely on that species, which can generate trivial or structurally decoupled contributions to the MAF. By excluding these cases, we focus on networks in which amplification arises from genuine inter-species structure rather than from embedded single-species self-coupling motifs.

For our numerical experiments, we choose the weight vector \(\w=\mbf{1}\), so that \(\epsilon=1\). The rescaled rate constants are therefore identical to the original rate constants. For each realization, we
sample them uniformly from \([0,1]\) and then normalize them by setting the largest rescaled rate constant to unity. Since the rescaled rate constants have dimensions of frequency, this normalization simply fixes the unit of
time.

\begin{figure*}[t]
    \centering
    
    \begin{subfigure}{0.47\textwidth}
        \centering \includegraphics[width=\linewidth]{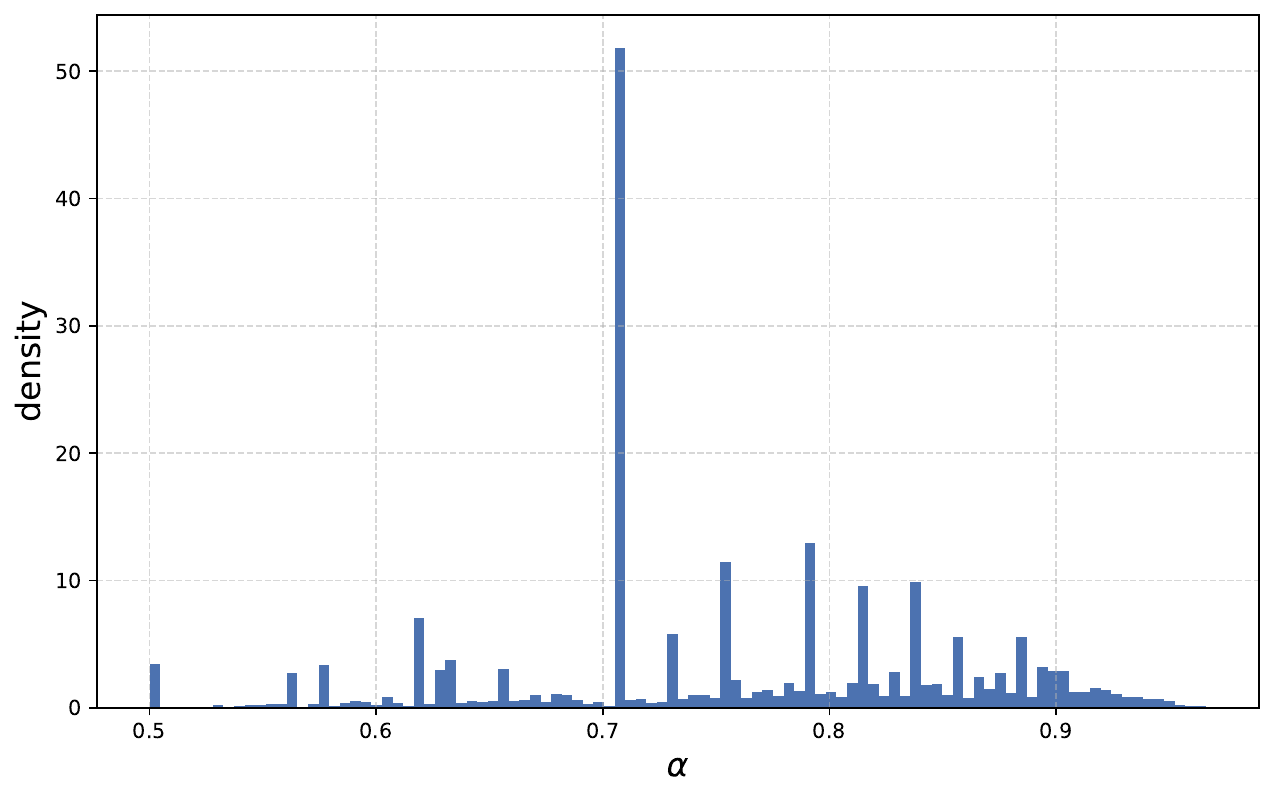}
    \end{subfigure}
    \hfill
    \begin{subfigure}{0.47\textwidth}
        \centering
        \includegraphics[width=\linewidth]{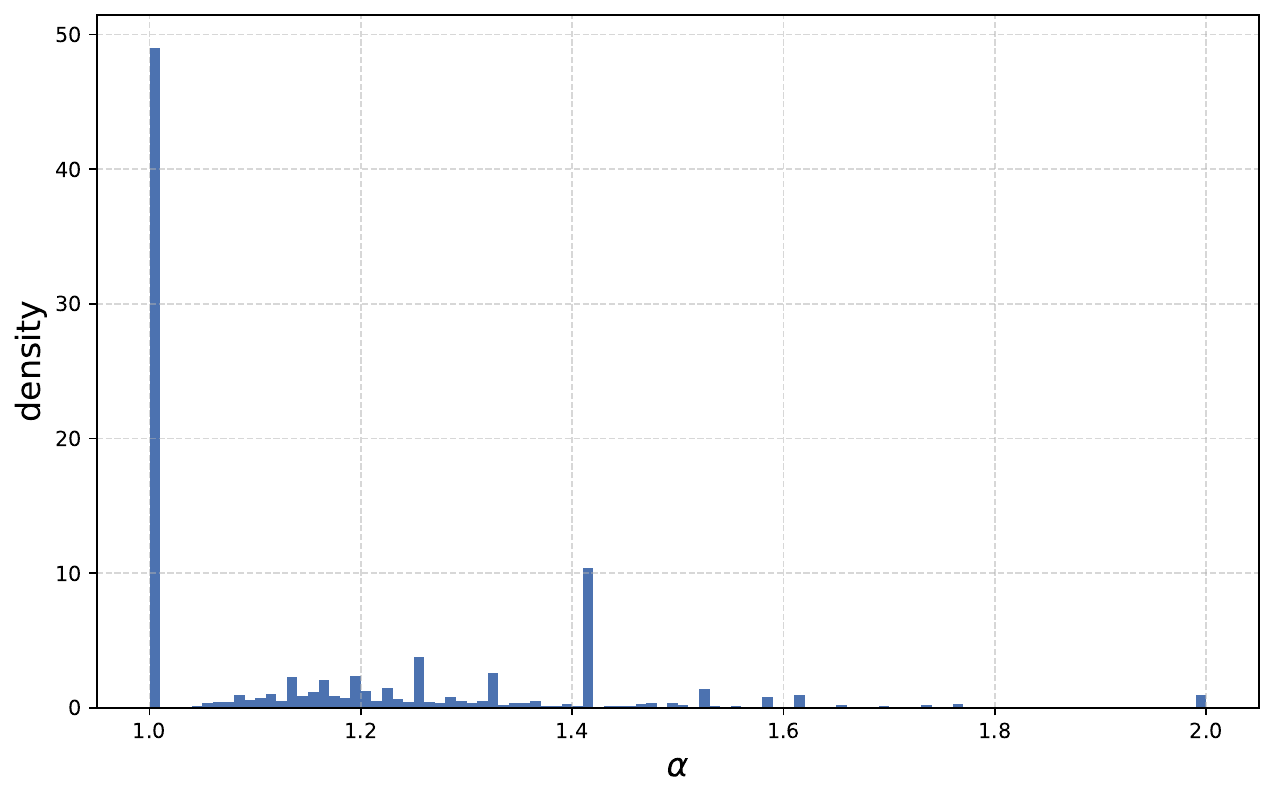}
    \end{subfigure} 
    \caption{Density of $\alpha$ values corresponding to an ensemble of random autonomous, unambiguous CRNs, similarly to Fig.~\ref{fig:core_labelled} of main text, except that now a full enumeration of all suitable networks up to size $4 \times 4$ has been performed. 
Left: autoinhibitory networks ($\alpha < 1$). 
Right: autocatalytic ($\alpha > 1$) and steady-state flux ($\alpha = 1$) networks. 
}  \label{fig:enumeration_figs}
\end{figure*}

For Figure 2 and 3 of the main text, we have generated $60000$ networks with $\alpha \geq 1$, and $40000$ networks with $\alpha<1$ satisfying the above conditions. 
For all these networks, we numerically computed their values of $\alpha$ using a linear programming code based on Refs. \cite{blanco2025identifyingselfamplifyinghypergraphstructures,sargent2019neumann}. Processing all these networks required a total of \(3.3\) CPU hours and \(2.4\) GPU hours to obtain these results on a server equipped with Ice Lake-SP processors.

The precise form of the density of \(\alpha\) depends on the sampling procedure. Although our large-scale random-network sampling exhaustively covers networks of small size, often with multiple samples of the smallest networks, it is not sufficient to sample all admissible random networks of size \(12\times 12\). To assess the effect of undersampling larger networks, we exactly enumerated all random networks up to size $4 \times 4$ satisfying the above conditions. The resulting distribution of \(\alpha\) is shown in Fig.~\ref{fig:enumeration_figs}. In this case, \(541{,}917\) random networks satisfy the conditions, and the enumeration required \(17\,\mathrm{h}\) of computation time. Although the peak amplitudes are affected by the reduced network size, the distribution retains several features of the main-text distribution, including prominent peaks at \(\alpha=\sqrt{2}\), \(\alpha=1\), and \(\alpha=1/\sqrt{2}\). The peak at \(\alpha=2\), however, is strongly reduced, indicating that many networks larger than \(4\times4\) contribute to this peak.

\subsection*{ E2. Bounds on the MAF}
By definition of the reaction order, we have 
\be
\forall \,r: \;\sum_s (\St^\pm)_s^r \leq M^\pm.
\ee
By the definition of the reaction order and the property of autonomy (every reaction consumes and produces at least one species), we obtain the right and left inequalities, respectively, in the following equation:
\be
 \forall \, r: \; 1 \leq \sum_s (\St^\pm)_s^r \leq M^\pm. 
 \label{eq:order_autonomy}
\ee
Consider the quantity
\be
Q(\x) := \frac{\sum_{s,r} (\Sp)_s^r \x_r }{\sum_{s,r} (\Sm)_s^r \x_r}.
\ee
By Eq.~\ref{eq:order_autonomy}, we have
\be
\sum_r \x_r \leq \sum_{s,r} (\St^\pm)_s^r \x_r \leq M^\pm \sum_r \x_r,
\ee
which implies
\be
\frac{1}{M^-} \leq Q(\x) \leq M^+. 
\label{eq:bound_Q}
\ee

Recall that the MAF \(\alpha\) is defined as the maximum over $\x > \0$ of the minimum ratios \((\mathbb{S}^+ \mathbf{x})_s / (\mathbb{S}^- \mathbf{x})_s\). 
Suppose $\alpha > M^+$, and there exists $\x$ such that \(\forall\, s:\, (\mathbb{S}^+ \mathbf{x})_s / (\mathbb{S}^- \mathbf{x})_s > M^+.\) 
Clearly,
\be
Q(\x) = \frac{\sum_s (\mathbb{S}^+ \mathbf{x})_s}{\sum_s (\Sm \mathbf{x})_s} > M^+,
\ee
which would violate the right-hand side of Eq.~\ref{eq:bound_Q} and yield a contradiction. Thus $\alpha \leq M^+$. 

Now suppose $\alpha < 1/M^-$ and there exists $\x^*$ such that 
\be
\min_s \frac{(\mathbb{S}^+ \x^*)_s}{(\mathbb{S}^- \x^*)_s} < \frac{1}{M^-}.
\ee
Following the earlier discussion around Eq.~\ref{eq:GEP} of the solution for $\alpha$, there is a set of species $\mcl{S}'$ with $|\mcl{S}'| \geq 1$ in an irreducible subnetwork of the CRN such that the minimum is attained simultaneously. Defining 
\be
Q'(\x) := \frac{\sum_{s \in \mcl{S}',r} (\Sp)_s^r \x_r }{\sum_{s \in \mcl{S}',r} (\Sm)_s^r \x_r},
\ee
and using the same arguments as before, we see that \(\forall \, \x:\, 1/M^- \leq Q'(\x)\). However, if $\alpha < 1/M^-$, then $Q'(\x^*) < 1/M^-$, which violates the inequality for $Q'(\x)$. Again we obtain a contradiction, and thus $\alpha \geq 1/M^-$. 

Putting all the previously stated results together, we obtain
\be
\frac{1}{M^-} \leq \alpha \leq M^+.
\ee

In what follows, we present examples of networks that saturate these bounds. For ambiguous networks with $|\mcl{S}| = 1$, the lower and upper bounds are saturated by the following networks, respectively:
\be
\{(M^-) X \to X\}
\quad \text{and} \quad
\{X \to (M^+) X\}.
\ee
An unambiguous network with $M^+ = 2$ and $|\mcl{S}| = 2$ that saturates the upper bound is (the construction for the lower bound is analogous):
\begin{align}
X_1 &\to 2 X_2,\\
X_2 &\to 2 X_1.
\end{align}
Similarly, an unambiguous network with $M^+ = 2$ and $|\mcl{S}| = 3$ that saturates the upper bound is:
\begin{align}
X_1 &\to X_2 + X_3,\\
X_2 &\to X_1 + X_3,\\
X_3 &\to X_1 + X_2.
\end{align}
Observe that the last network is exactly the type~V autocatalytic core.

\subsection{ E3. $\tilde{\Lambda}$ vs $\delta$ for $M^+=M^-=10$}

The figure below shows that the spectrum of values of $\alpha$ becomes continuous in the limit that $M^+$ and $M^-$ become large.

One can understand this observation qualitatively as follows. First, as explained on page 2 of the main text, the MAF is determined by the irreducible autocatalytic subnetwork with the largest MAF. This subnetwork need not be a minimal autocatalytic subnetwork, or core. Second, the classification into five core types only applies to reversible networks. For the irreversible networks considered here, 
we have not performed an exact enumeration, but we know that this number is larger than 5. 
Moreover, even within a given topological core type, branches can be extended to produce networks that belong to the same type but have different MAFs. For example, the networks
\begin{equation*}
    \{A \to B,\; B \to 2A\}\; \text{ and }\;
    \{A \to B,\; B \to C,\; C \to 2A\}
\end{equation*}
are both type-1 cores, but have different MAF values, namely
\(\alpha \simeq 1.41\) and \(\alpha \simeq 1.26\), respectively.\\
Our figures suggest the following picture. 
For \(M^+=M^-=2\), 
the presence of more than five core types in irreversible networks can explain the additional peaks in the distribution, while variants within a given core type can account for the slight broadening around peaks associated with known reversible core types.\\
For networks with \(M^+=M^-=10\), the irreducible autocatalytic subnetworks that determine \(\alpha\) appear to consist of multiple coupled cores. This may explain the broader spectrum observed in this case.

\begin{figure*}[t]
    \centering
    
    \begin{subfigure}{0.45\textwidth}
        \centering
        \includegraphics[width=\linewidth]{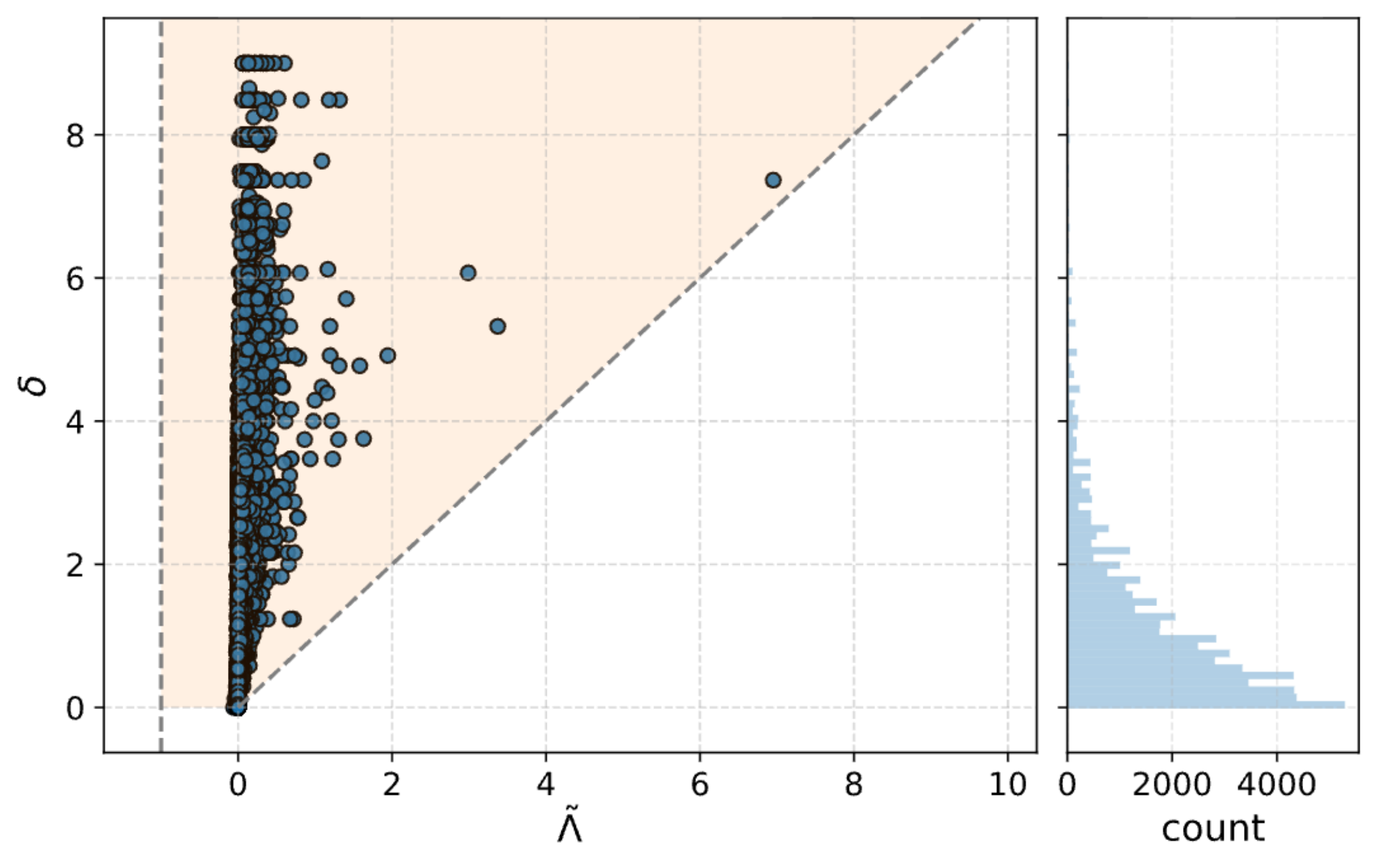}
    \end{subfigure}
    \hfill
    \begin{subfigure}{0.45\textwidth}
        \centering
        \includegraphics[width=\linewidth]{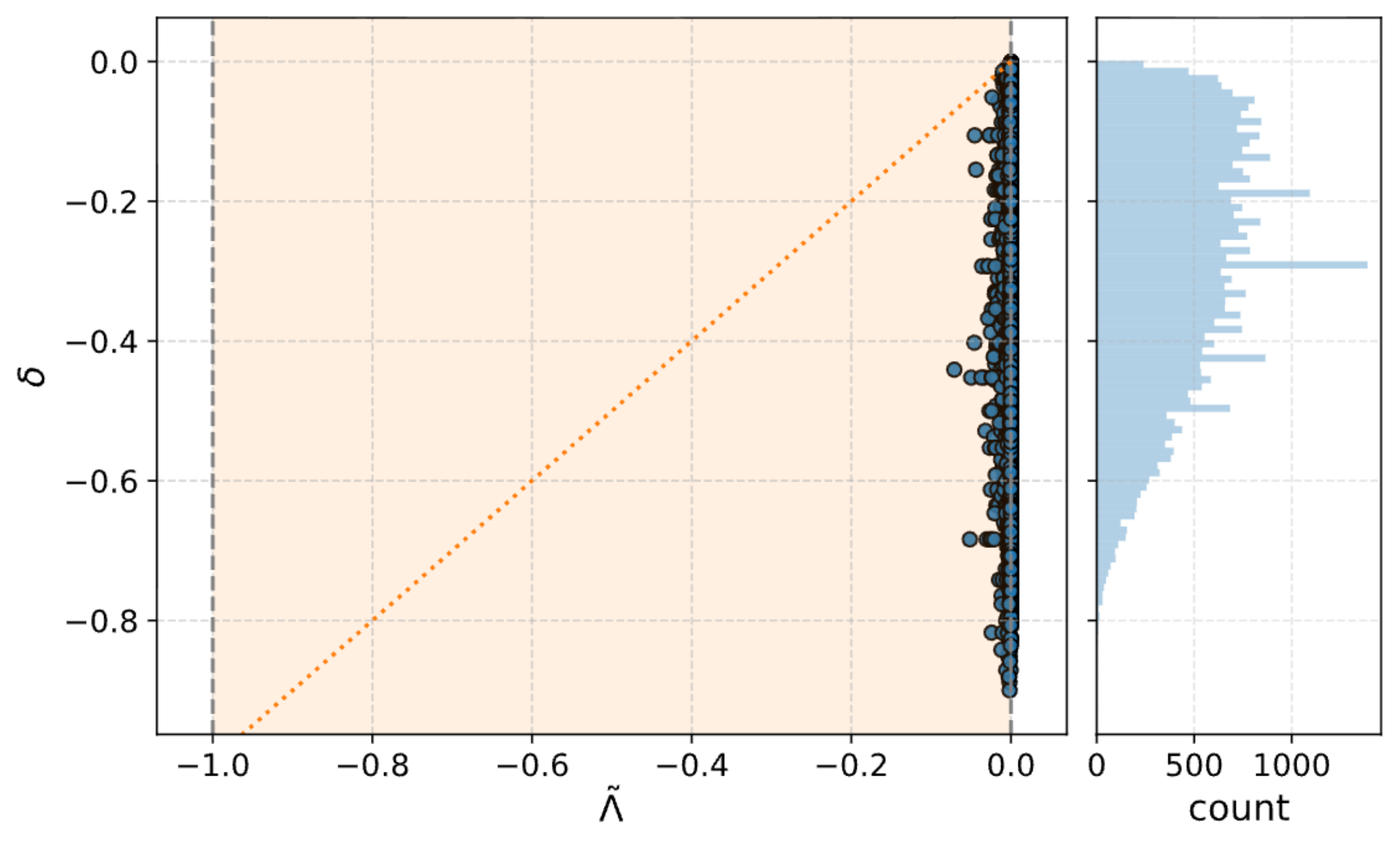}
    \end{subfigure}
    
    \caption{Growth parameter $\delta=\alpha-1$ versus normalized growth rate 
$\tilde{\Lambda}=\Lambda/\|\Sm\|_{\bm{\kappa}}$ for random autonomous, unambiguous CRNs with 
$2 \leq |\mcl{S}|,|\mcl{R}| \leq 12$ with $M^+=M^-=10$. 
Left: autocatalytic networks ($\alpha > 1$, $\delta > 0$) and networks which admit a steady-state flux ($\alpha = 1$, $\delta = 0$). 
Right: autoinhibitory networks ($\alpha < 1$, $\delta < 0$). 
}
\label{fig:shrinkgrow_M10}
\end{figure*}

\begin{thebibliography}{28}%
\makeatletter
\providecommand \@ifxundefined [1]{%
 \@ifx{#1\undefined}
}%
\providecommand \@ifnum [1]{%
 \ifnum #1\expandafter \@firstoftwo
 \else \expandafter \@secondoftwo
 \fi
}%
\providecommand \@ifx [1]{%
 \ifx #1\expandafter \@firstoftwo
 \else \expandafter \@secondoftwo
 \fi
}%
\providecommand \natexlab [1]{#1}%
\providecommand \enquote  [1]{``#1''}%
\providecommand \bibnamefont  [1]{#1}%
\providecommand \bibfnamefont [1]{#1}%
\providecommand \citenamefont [1]{#1}%
\providecommand \href@noop [0]{\@secondoftwo}%
\providecommand \href [0]{\begingroup \@sanitize@url \@href}%
\providecommand \@href[1]{\@@startlink{#1}\@@href}%
\providecommand \@@href[1]{\endgroup#1\@@endlink}%
\providecommand \@sanitize@url [0]{\catcode `\\12\catcode `\$12\catcode `\&12\catcode `\#12\catcode `\^12\catcode `\_12\catcode `\%12\relax}%
\providecommand \@@startlink[1]{}%
\providecommand \@@endlink[0]{}%
\providecommand \url  [0]{\begingroup\@sanitize@url \@url }%
\providecommand \@url [1]{\endgroup\@href {#1}{\urlprefix }}%
\providecommand \urlprefix  [0]{URL }%
\providecommand \Eprint [0]{\href }%
\providecommand \doibase [0]{https://doi.org/}%
\providecommand \selectlanguage [0]{\@gobble}%
\providecommand \bibinfo  [0]{\@secondoftwo}%
\providecommand \bibfield  [0]{\@secondoftwo}%
\providecommand \translation [1]{[#1]}%
\providecommand \BibitemOpen [0]{}%
\providecommand \bibitemStop [0]{}%
\providecommand \bibitemNoStop [0]{.\EOS\space}%
\providecommand \EOS [0]{\spacefactor3000\relax}%
\providecommand \BibitemShut  [1]{\csname bibitem#1\endcsname}%
\let\auto@bib@innerbib\@empty
\bibitem [{\citenamefont {Eigen}\ and\ \citenamefont {Schuster}(1977)}]{eigen1977principle}%
  \BibitemOpen
  \bibfield  {author} {\bibinfo {author} {\bibfnamefont {M.}~\bibnamefont {Eigen}}\ and\ \bibinfo {author} {\bibfnamefont {P.}~\bibnamefont {Schuster}},\ }\href@noop {} {\bibfield  {journal} {\bibinfo  {journal} {Naturwissenschaften}\ }\textbf {\bibinfo {volume} {64}},\ \bibinfo {pages} {541} (\bibinfo {year} {1977})}\BibitemShut {NoStop}%
\bibitem [{\citenamefont {Baum}\ \emph {et~al.}(2023)\citenamefont {Baum}, \citenamefont {Peng}, \citenamefont {Dolson}, \citenamefont {Smith}, \citenamefont {Plum},\ and\ \citenamefont {Gagrani}}]{baum2023ecology}%
  \BibitemOpen
  \bibfield  {author} {\bibinfo {author} {\bibfnamefont {D.~A.}\ \bibnamefont {Baum}}, \bibinfo {author} {\bibfnamefont {Z.}~\bibnamefont {Peng}}, \bibinfo {author} {\bibfnamefont {E.}~\bibnamefont {Dolson}}, \bibinfo {author} {\bibfnamefont {E.}~\bibnamefont {Smith}}, \bibinfo {author} {\bibfnamefont {A.~M.}\ \bibnamefont {Plum}},\ and\ \bibinfo {author} {\bibfnamefont {P.}~\bibnamefont {Gagrani}},\ }\href@noop {} {\bibfield  {journal} {\bibinfo  {journal} {Journal of the Royal Society Interface}\ }\textbf {\bibinfo {volume} {20}} (\bibinfo {year} {2023})}\BibitemShut {NoStop}%
\bibitem [{\citenamefont {Ameta}\ \emph {et~al.}(2021)\citenamefont {Ameta}, \citenamefont {Matsubara}, \citenamefont {Chakraborty}, \citenamefont {Krishna},\ and\ \citenamefont {Thutupalli}}]{ameta_self-reproduction_2021}%
  \BibitemOpen
  \bibfield  {author} {\bibinfo {author} {\bibfnamefont {S.}~\bibnamefont {Ameta}}, \bibinfo {author} {\bibfnamefont {Y.~J.}\ \bibnamefont {Matsubara}}, \bibinfo {author} {\bibfnamefont {N.}~\bibnamefont {Chakraborty}}, \bibinfo {author} {\bibfnamefont {S.}~\bibnamefont {Krishna}},\ and\ \bibinfo {author} {\bibfnamefont {S.}~\bibnamefont {Thutupalli}},\ }\href@noop {} {\bibfield  {journal} {\bibinfo  {journal} {Life}\ }\textbf {\bibinfo {volume} {11}},\ \bibinfo {pages} {308} (\bibinfo {year} {2021})}\BibitemShut {NoStop}%
\bibitem [{\citenamefont {Blokhuis}\ \emph {et~al.}(2020)\citenamefont {Blokhuis}, \citenamefont {Lacoste},\ and\ \citenamefont {Nghe}}]{blokhuis2020universal}%
  \BibitemOpen
  \bibfield  {author} {\bibinfo {author} {\bibfnamefont {A.}~\bibnamefont {Blokhuis}}, \bibinfo {author} {\bibfnamefont {D.}~\bibnamefont {Lacoste}},\ and\ \bibinfo {author} {\bibfnamefont {P.}~\bibnamefont {Nghe}},\ }\href@noop {} {\bibfield  {journal} {\bibinfo  {journal} {Proceedings of the National Academy of Sciences}\ }\textbf {\bibinfo {volume} {117}},\ \bibinfo {pages} {25230} (\bibinfo {year} {2020})}\BibitemShut {NoStop}%
\bibitem [{\citenamefont {Gagrani}\ \emph {et~al.}(2024)\citenamefont {Gagrani}, \citenamefont {Blanco}, \citenamefont {Smith},\ and\ \citenamefont {Baum}}]{gagrani2024polyhedral}%
  \BibitemOpen
  \bibfield  {author} {\bibinfo {author} {\bibfnamefont {P.}~\bibnamefont {Gagrani}}, \bibinfo {author} {\bibfnamefont {V.}~\bibnamefont {Blanco}}, \bibinfo {author} {\bibfnamefont {E.}~\bibnamefont {Smith}},\ and\ \bibinfo {author} {\bibfnamefont {D.}~\bibnamefont {Baum}},\ }\href@noop {} {\bibfield  {journal} {\bibinfo  {journal} {Journal of Mathematical Chemistry}\ }\textbf {\bibinfo {volume} {62}},\ \bibinfo {pages} {1012} (\bibinfo {year} {2024})}\BibitemShut {NoStop}%
\bibitem [{\citenamefont {Golnik}\ \emph {et~al.}(2025)\citenamefont {Golnik}, \citenamefont {Gatter}, \citenamefont {Stadler},\ and\ \citenamefont {Vassena}}]{golnik2025enumeration}%
  \BibitemOpen
  \bibfield  {author} {\bibinfo {author} {\bibfnamefont {R.}~\bibnamefont {Golnik}}, \bibinfo {author} {\bibfnamefont {T.}~\bibnamefont {Gatter}}, \bibinfo {author} {\bibfnamefont {P.~F.}\ \bibnamefont {Stadler}},\ and\ \bibinfo {author} {\bibfnamefont {N.}~\bibnamefont {Vassena}},\ }\href@noop {} {\bibfield  {journal} {\bibinfo  {journal} {arXiv preprint arXiv:2511.18883}\ } (\bibinfo {year} {2025})}\BibitemShut {NoStop}%
\bibitem [{\citenamefont {Andersen}\ \emph {et~al.}(2020)\citenamefont {Andersen}, \citenamefont {Flamm}, \citenamefont {Merkle},\ and\ \citenamefont {Stadler}}]{andersen_defining_2020}%
  \BibitemOpen
  \bibfield  {author} {\bibinfo {author} {\bibfnamefont {J.~L.}\ \bibnamefont {Andersen}}, \bibinfo {author} {\bibfnamefont {C.}~\bibnamefont {Flamm}}, \bibinfo {author} {\bibfnamefont {D.}~\bibnamefont {Merkle}},\ and\ \bibinfo {author} {\bibfnamefont {P.~F.}\ \bibnamefont {Stadler}},\ }\href@noop {} {\bibfield  {journal} {\bibinfo  {journal} {Journal of Systems Chemistry}\ }\textbf {\bibinfo {volume} {8}},\ \bibinfo {pages} {121} (\bibinfo {year} {2020})}\BibitemShut {NoStop}%
\bibitem [{\citenamefont {Fang}\ \emph {et~al.}(2020)\citenamefont {Fang}, \citenamefont {Lloyd},\ and\ \citenamefont {Palsson}}]{fang_reconstructing_2020}%
  \BibitemOpen
  \bibfield  {author} {\bibinfo {author} {\bibfnamefont {X.}~\bibnamefont {Fang}}, \bibinfo {author} {\bibfnamefont {C.~J.}\ \bibnamefont {Lloyd}},\ and\ \bibinfo {author} {\bibfnamefont {B.~O.}\ \bibnamefont {Palsson}},\ }\href {https://doi.org/10.1038/s41579-020-00440-4} {\bibfield  {journal} {\bibinfo  {journal} {Nat Rev Microbiol}\ }\textbf {\bibinfo {volume} {18}},\ \bibinfo {pages} {731} (\bibinfo {year} {2020})}\BibitemShut {NoStop}%
\bibitem [{\citenamefont {Lin}\ \emph {et~al.}(2020)\citenamefont {Lin}, \citenamefont {Kussell}, \citenamefont {Young},\ and\ \citenamefont {Jacobs-Wagner}}]{lin2020origin}%
  \BibitemOpen
  \bibfield  {author} {\bibinfo {author} {\bibfnamefont {W.-H.}\ \bibnamefont {Lin}}, \bibinfo {author} {\bibfnamefont {E.}~\bibnamefont {Kussell}}, \bibinfo {author} {\bibfnamefont {L.-S.}\ \bibnamefont {Young}},\ and\ \bibinfo {author} {\bibfnamefont {C.}~\bibnamefont {Jacobs-Wagner}},\ }\href@noop {} {\bibfield  {journal} {\bibinfo  {journal} {Proceedings of the National Academy of Sciences}\ }\textbf {\bibinfo {volume} {117}},\ \bibinfo {pages} {27795} (\bibinfo {year} {2020})}\BibitemShut {NoStop}%
\bibitem [{\citenamefont {Singh}\ and\ \citenamefont {Jain}(2023)}]{singh2023multistable}%
  \BibitemOpen
  \bibfield  {author} {\bibinfo {author} {\bibfnamefont {A.~Y.}\ \bibnamefont {Singh}}\ and\ \bibinfo {author} {\bibfnamefont {S.}~\bibnamefont {Jain}},\ }\href@noop {} {\bibfield  {journal} {\bibinfo  {journal} {Life}\ }\textbf {\bibinfo {volume} {13}},\ \bibinfo {pages} {2327} (\bibinfo {year} {2023})}\BibitemShut {NoStop}%
\bibitem [{\citenamefont {Dourado}\ and\ \citenamefont {Lercher}(2020)}]{dourado_analytical_2020}%
  \BibitemOpen
  \bibfield  {author} {\bibinfo {author} {\bibfnamefont {H.}~\bibnamefont {Dourado}}\ and\ \bibinfo {author} {\bibfnamefont {M.~J.}\ \bibnamefont {Lercher}},\ }\href@noop {} {\bibfield  {journal} {\bibinfo  {journal} {Nat Commun}\ }\textbf {\bibinfo {volume} {11}},\ \bibinfo {pages} {1226} (\bibinfo {year} {2020})}\BibitemShut {NoStop}%
\bibitem [{\citenamefont {Pandey}\ \emph {et~al.}(2020)\citenamefont {Pandey}, \citenamefont {Singh},\ and\ \citenamefont {Jain}}]{pandey2020exponential}%
  \BibitemOpen
  \bibfield  {author} {\bibinfo {author} {\bibfnamefont {P.~P.}\ \bibnamefont {Pandey}}, \bibinfo {author} {\bibfnamefont {H.}~\bibnamefont {Singh}},\ and\ \bibinfo {author} {\bibfnamefont {S.}~\bibnamefont {Jain}},\ }\href@noop {} {\bibfield  {journal} {\bibinfo  {journal} {Physical Review E}\ }\textbf {\bibinfo {volume} {101}},\ \bibinfo {pages} {062406} (\bibinfo {year} {2020})}\BibitemShut {NoStop}%
\bibitem [{\citenamefont {Liebermeister}\ \emph {et~al.}(2010)\citenamefont {Liebermeister}, \citenamefont {Uhlendorf},\ and\ \citenamefont {Klipp}}]{Liebermeister2010}%
  \BibitemOpen
  \bibfield  {author} {\bibinfo {author} {\bibfnamefont {W.}~\bibnamefont {Liebermeister}}, \bibinfo {author} {\bibfnamefont {J.}~\bibnamefont {Uhlendorf}},\ and\ \bibinfo {author} {\bibfnamefont {E.}~\bibnamefont {Klipp}},\ }\href@noop {} {\bibfield  {journal} {\bibinfo  {journal} {Bioinformatics}\ }\textbf {\bibinfo {volume} {26}},\ \bibinfo {pages} {1528} (\bibinfo {year} {2010})}\BibitemShut {NoStop}%
\bibitem [{\citenamefont {Pandey}\ and\ \citenamefont {Jain}(2016)}]{pandey2016analytic}%
  \BibitemOpen
  \bibfield  {author} {\bibinfo {author} {\bibfnamefont {P.~P.}\ \bibnamefont {Pandey}}\ and\ \bibinfo {author} {\bibfnamefont {S.}~\bibnamefont {Jain}},\ }\href@noop {} {\bibfield  {journal} {\bibinfo  {journal} {Theory in Biosciences}\ }\textbf {\bibinfo {volume} {135}},\ \bibinfo {pages} {121} (\bibinfo {year} {2016})}\BibitemShut {NoStop}%
\bibitem [{\citenamefont {M{\"u}ller}\ \emph {et~al.}(2022)\citenamefont {M{\"u}ller}, \citenamefont {Sz{\'e}liov{\'a}},\ and\ \citenamefont {Zanghellini}}]{muller2022elementary}%
  \BibitemOpen
  \bibfield  {author} {\bibinfo {author} {\bibfnamefont {S.}~\bibnamefont {M{\"u}ller}}, \bibinfo {author} {\bibfnamefont {D.}~\bibnamefont {Sz{\'e}liov{\'a}}},\ and\ \bibinfo {author} {\bibfnamefont {J.}~\bibnamefont {Zanghellini}},\ }\href@noop {} {\bibfield  {journal} {\bibinfo  {journal} {PLoS computational biology}\ }\textbf {\bibinfo {volume} {18}},\ \bibinfo {pages} {e1009843} (\bibinfo {year} {2022})}\BibitemShut {NoStop}%
\bibitem [{\citenamefont {Neumann}(1945)}]{neumann1945model}%
  \BibitemOpen
  \bibfield  {author} {\bibinfo {author} {\bibfnamefont {J.~v.}\ \bibnamefont {Neumann}},\ }\href@noop {} {\bibfield  {journal} {\bibinfo  {journal} {The Review of Economic Studies}\ }\textbf {\bibinfo {volume} {13}},\ \bibinfo {pages} {1} (\bibinfo {year} {1945})}\BibitemShut {NoStop}%
\bibitem [{\citenamefont {Sargent}\ and\ \citenamefont {Stachurski}(2019)}]{sargent2019neumann}%
  \BibitemOpen
  \bibfield  {author} {\bibinfo {author} {\bibfnamefont {T.~J.}\ \bibnamefont {Sargent}}\ and\ \bibinfo {author} {\bibfnamefont {J.}~\bibnamefont {Stachurski}},\ }\href@noop {} {\bibfield  {journal} {\bibinfo  {journal} {Online Textbook}\ } (\bibinfo {year} {2019})}\BibitemShut {NoStop}%
\bibitem [{\citenamefont {Gale}(1989)}]{gale1989theory}%
  \BibitemOpen
  \bibfield  {author} {\bibinfo {author} {\bibfnamefont {D.}~\bibnamefont {Gale}},\ }\href@noop {} {\emph {\bibinfo {title} {The theory of linear economic models}}}\ (\bibinfo  {publisher} {University of Chicago press},\ \bibinfo {year} {1989})\BibitemShut {NoStop}%
\bibitem [{\citenamefont {Blanco}\ \emph {et~al.}(2025)\citenamefont {Blanco}, \citenamefont {González},\ and\ \citenamefont {Gagrani}}]{blanco2025identifyingselfamplifyinghypergraphstructures}%
  \BibitemOpen
  \bibfield  {author} {\bibinfo {author} {\bibfnamefont {V.}~\bibnamefont {Blanco}}, \bibinfo {author} {\bibfnamefont {G.}~\bibnamefont {González}},\ and\ \bibinfo {author} {\bibfnamefont {P.}~\bibnamefont {Gagrani}},\ }\href {https://arxiv.org/abs/2412.15776} {\bibinfo {title} {Identifying self-amplifying hypergraph structures through mathematical optimization}} (\bibinfo {year} {2025}),\ \Eprint {https://arxiv.org/abs/2412.15776} {arXiv:2412.15776 [math.OC]} \BibitemShut {NoStop}%
\bibitem [{\citenamefont {De~Martino}\ \emph {et~al.}(2012)\citenamefont {De~Martino}, \citenamefont {Marinari},\ and\ \citenamefont {Romualdi}}]{de_martino_von_2012}%
  \BibitemOpen
  \bibfield  {author} {\bibinfo {author} {\bibfnamefont {A.}~\bibnamefont {De~Martino}}, \bibinfo {author} {\bibfnamefont {E.}~\bibnamefont {Marinari}},\ and\ \bibinfo {author} {\bibfnamefont {A.}~\bibnamefont {Romualdi}},\ }\href {https://doi.org/10.1140/epjst/e2012-01653-8} {\bibfield  {journal} {\bibinfo  {journal} {Eur. Phys. J. Spec. Top.}\ }\textbf {\bibinfo {volume} {212}},\ \bibinfo {pages} {45} (\bibinfo {year} {2012})}\BibitemShut {NoStop}%
\bibitem [{\citenamefont {Qian}\ and\ \citenamefont {Bishop}(2010)}]{qian_chemical_2010}%
  \BibitemOpen
  \bibfield  {author} {\bibinfo {author} {\bibfnamefont {H.}~\bibnamefont {Qian}}\ and\ \bibinfo {author} {\bibfnamefont {L.~M.}\ \bibnamefont {Bishop}},\ }\href@noop {} {\bibfield  {journal} {\bibinfo  {journal} {Int J Mol Sci}\ }\textbf {\bibinfo {volume} {11}},\ \bibinfo {pages} {3472} (\bibinfo {year} {2010})}\BibitemShut {NoStop}%
\bibitem [{\citenamefont {Sol{\'e}}\ \emph {et~al.}(2024)\citenamefont {Sol{\'e}}, \citenamefont {Kempes}, \citenamefont {Corominas-Murtra}, \citenamefont {De~Domenico}, \citenamefont {Kolchinsky}, \citenamefont {Lachmann}, \citenamefont {Libby}, \citenamefont {Saavedra}, \citenamefont {Smith},\ and\ \citenamefont {Wolpert}}]{sole_fundamental_2024}%
  \BibitemOpen
  \bibfield  {author} {\bibinfo {author} {\bibfnamefont {R.}~\bibnamefont {Sol{\'e}}}, \bibinfo {author} {\bibfnamefont {C.~P.}\ \bibnamefont {Kempes}}, \bibinfo {author} {\bibfnamefont {B.}~\bibnamefont {Corominas-Murtra}}, \bibinfo {author} {\bibfnamefont {M.}~\bibnamefont {De~Domenico}}, \bibinfo {author} {\bibfnamefont {A.}~\bibnamefont {Kolchinsky}}, \bibinfo {author} {\bibfnamefont {M.}~\bibnamefont {Lachmann}}, \bibinfo {author} {\bibfnamefont {E.}~\bibnamefont {Libby}}, \bibinfo {author} {\bibfnamefont {S.}~\bibnamefont {Saavedra}}, \bibinfo {author} {\bibfnamefont {E.}~\bibnamefont {Smith}},\ and\ \bibinfo {author} {\bibfnamefont {D.}~\bibnamefont {Wolpert}},\ }\href@noop {} {\bibfield  {journal} {\bibinfo  {journal} {Interface Focus}\ }\textbf {\bibinfo {volume} {14}},\ \bibinfo {pages} {20240010} (\bibinfo {year} {2024})}\BibitemShut {NoStop}%
\bibitem [{\citenamefont {Jain}\ and\ \citenamefont {Krishna}(2001)}]{jain_model_2001}%
  \BibitemOpen
  \bibfield  {author} {\bibinfo {author} {\bibfnamefont {S.}~\bibnamefont {Jain}}\ and\ \bibinfo {author} {\bibfnamefont {S.}~\bibnamefont {Krishna}},\ }\href@noop {} {\bibfield  {journal} {\bibinfo  {journal} {Proceedings of the National Academy of Sciences}\ }\textbf {\bibinfo {volume} {98}},\ \bibinfo {pages} {543} (\bibinfo {year} {2001})}\BibitemShut {NoStop}%
\bibitem [{\citenamefont {Sarkar}\ and\ \citenamefont {England}(2019)}]{sarkar_design_2019}%
  \BibitemOpen
  \bibfield  {author} {\bibinfo {author} {\bibfnamefont {S.}~\bibnamefont {Sarkar}}\ and\ \bibinfo {author} {\bibfnamefont {J.~L.}\ \bibnamefont {England}},\ }\href@noop {} {\bibfield  {journal} {\bibinfo  {journal} {Phys. Rev. E}\ }\textbf {\bibinfo {volume} {100}},\ \bibinfo {pages} {022414} (\bibinfo {year} {2019})}\BibitemShut {NoStop}%
\bibitem [{\citenamefont {Despons}\ \emph {et~al.}(2025)\citenamefont {Despons}, \citenamefont {Unterberger},\ and\ \citenamefont {Lacoste}}]{despons_stability_2025}%
  \BibitemOpen
  \bibfield  {author} {\bibinfo {author} {\bibfnamefont {A.}~\bibnamefont {Despons}}, \bibinfo {author} {\bibfnamefont {J.}~\bibnamefont {Unterberger}},\ and\ \bibinfo {author} {\bibfnamefont {D.}~\bibnamefont {Lacoste}},\ }\href {https://doi.org/10.1088/1367-2630/ae2153} {\bibfield  {journal} {\bibinfo  {journal} {New J. Phys.}\ }\textbf {\bibinfo {volume} {27}},\ \bibinfo {pages} {124201} (\bibinfo {year} {2025})},\ \bibinfo {note} {publisher: IOP Publishing}\BibitemShut {NoStop}%
\bibitem [{\citenamefont {Marehalli~Srinivas}\ \emph {et~al.}(2024)\citenamefont {Marehalli~Srinivas}, \citenamefont {Avanzini},\ and\ \citenamefont {Esposito}}]{marehalli_srinivas_thermodynamics_2024}%
  \BibitemOpen
  \bibfield  {author} {\bibinfo {author} {\bibfnamefont {S.~G.}\ \bibnamefont {Marehalli~Srinivas}}, \bibinfo {author} {\bibfnamefont {F.}~\bibnamefont {Avanzini}},\ and\ \bibinfo {author} {\bibfnamefont {M.}~\bibnamefont {Esposito}},\ }\href@noop {} {\bibfield  {journal} {\bibinfo  {journal} {Phys. Rev. Lett.}\ }\textbf {\bibinfo {volume} {132}},\ \bibinfo {pages} {268001} (\bibinfo {year} {2024})}\BibitemShut {NoStop}%
\bibitem [{\citenamefont {Kolchinsky}(2025)}]{kolchinsky2025thermodynamics}%
  \BibitemOpen
  \bibfield  {author} {\bibinfo {author} {\bibfnamefont {A.}~\bibnamefont {Kolchinsky}},\ }\href@noop {} {\bibfield  {journal} {\bibinfo  {journal} {Philosophical Transactions of the Royal Society B: Biological Sciences}\ }\textbf {\bibinfo {volume} {380}} (\bibinfo {year} {2025})}\BibitemShut {NoStop}%
\bibitem [{\citenamefont {Lazarescu}\ \emph {et~al.}(2019)\citenamefont {Lazarescu}, \citenamefont {Cossetto}, \citenamefont {Falasco},\ and\ \citenamefont {Esposito}}]{lazarescu_large_2019}%
  \BibitemOpen
  \bibfield  {author} {\bibinfo {author} {\bibfnamefont {A.}~\bibnamefont {Lazarescu}}, \bibinfo {author} {\bibfnamefont {T.}~\bibnamefont {Cossetto}}, \bibinfo {author} {\bibfnamefont {G.}~\bibnamefont {Falasco}},\ and\ \bibinfo {author} {\bibfnamefont {M.}~\bibnamefont {Esposito}},\ }\href@noop {} {\bibfield  {journal} {\bibinfo  {journal} {J. Chem. Phys.}\ }\textbf {\bibinfo {volume} {151}},\ \bibinfo {pages} {064117} (\bibinfo {year} {2019})}\BibitemShut {NoStop}%
\end{thebibliography}%


\begin{thebibliography}{8}%
\makeatletter
\providecommand \@ifxundefined [1]{%
 \@ifx{#1\undefined}
}%
\providecommand \@ifnum [1]{%
 \ifnum #1\expandafter \@firstoftwo
 \else \expandafter \@secondoftwo
 \fi
}%
\providecommand \@ifx [1]{%
 \ifx #1\expandafter \@firstoftwo
 \else \expandafter \@secondoftwo
 \fi
}%
\providecommand \natexlab [1]{#1}%
\providecommand \enquote  [1]{``#1''}%
\providecommand \bibnamefont  [1]{#1}%
\providecommand \bibfnamefont [1]{#1}%
\providecommand \citenamefont [1]{#1}%
\providecommand \href@noop [0]{\@secondoftwo}%
\providecommand \href [0]{\begingroup \@sanitize@url \@href}%
\providecommand \@href[1]{\@@startlink{#1}\@@href}%
\providecommand \@@href[1]{\endgroup#1\@@endlink}%
\providecommand \@sanitize@url [0]{\catcode `\\12\catcode `\$12\catcode `\&12\catcode `\#12\catcode `\^12\catcode `\_12\catcode `\%12\relax}%
\providecommand \@@startlink[1]{}%
\providecommand \@@endlink[0]{}%
\providecommand \url  [0]{\begingroup\@sanitize@url \@url }%
\providecommand \@url [1]{\endgroup\@href {#1}{\urlprefix }}%
\providecommand \urlprefix  [0]{URL }%
\providecommand \Eprint [0]{\href }%
\providecommand \doibase [0]{https://doi.org/}%
\providecommand \selectlanguage [0]{\@gobble}%
\providecommand \bibinfo  [0]{\@secondoftwo}%
\providecommand \bibfield  [0]{\@secondoftwo}%
\providecommand \translation [1]{[#1]}%
\providecommand \BibitemOpen [0]{}%
\providecommand \bibitemStop [0]{}%
\providecommand \bibitemNoStop [0]{.\EOS\space}%
\providecommand \EOS [0]{\spacefactor3000\relax}%
\providecommand \BibitemShut  [1]{\csname bibitem#1\endcsname}%
\let\auto@bib@innerbib\@empty
\bibitem [{\citenamefont {Gagrani}\ \emph {et~al.}(2024)\citenamefont {Gagrani}, \citenamefont {Blanco}, \citenamefont {Smith},\ and\ \citenamefont {Baum}}]{gagrani2024polyhedral}%
  \BibitemOpen
  \bibfield  {author} {\bibinfo {author} {\bibfnamefont {P.}~\bibnamefont {Gagrani}}, \bibinfo {author} {\bibfnamefont {V.}~\bibnamefont {Blanco}}, \bibinfo {author} {\bibfnamefont {E.}~\bibnamefont {Smith}},\ and\ \bibinfo {author} {\bibfnamefont {D.}~\bibnamefont {Baum}},\ }\href@noop {} {\bibfield  {journal} {\bibinfo  {journal} {Journal of Mathematical Chemistry}\ }\textbf {\bibinfo {volume} {62}},\ \bibinfo {pages} {1012} (\bibinfo {year} {2024})}\BibitemShut {NoStop}%
\bibitem [{\citenamefont {Blanco}\ \emph {et~al.}(2025)\citenamefont {Blanco}, \citenamefont {González},\ and\ \citenamefont {Gagrani}}]{blanco2025identifyingselfamplifyinghypergraphstructures}%
  \BibitemOpen
  \bibfield  {author} {\bibinfo {author} {\bibfnamefont {V.}~\bibnamefont {Blanco}}, \bibinfo {author} {\bibfnamefont {G.}~\bibnamefont {González}},\ and\ \bibinfo {author} {\bibfnamefont {P.}~\bibnamefont {Gagrani}},\ }\href {https://arxiv.org/abs/2412.15776} {\bibinfo {title} {Identifying self-amplifying hypergraph structures through mathematical optimization}} (\bibinfo {year} {2025}),\ \Eprint {https://arxiv.org/abs/2412.15776} {arXiv:2412.15776 [math.OC]} \BibitemShut {NoStop}%
\bibitem [{\citenamefont {Blokhuis}\ \emph {et~al.}(2020)\citenamefont {Blokhuis}, \citenamefont {Lacoste},\ and\ \citenamefont {Nghe}}]{blokhuis2020universal}%
  \BibitemOpen
  \bibfield  {author} {\bibinfo {author} {\bibfnamefont {A.}~\bibnamefont {Blokhuis}}, \bibinfo {author} {\bibfnamefont {D.}~\bibnamefont {Lacoste}},\ and\ \bibinfo {author} {\bibfnamefont {P.}~\bibnamefont {Nghe}},\ }\href@noop {} {\bibfield  {journal} {\bibinfo  {journal} {Proceedings of the National Academy of Sciences}\ }\textbf {\bibinfo {volume} {117}},\ \bibinfo {pages} {25230} (\bibinfo {year} {2020})}\BibitemShut {NoStop}%
\bibitem [{\citenamefont {Gale}(1989)}]{gale1989theory}%
  \BibitemOpen
  \bibfield  {author} {\bibinfo {author} {\bibfnamefont {D.}~\bibnamefont {Gale}},\ }\href@noop {} {\emph {\bibinfo {title} {The theory of linear economic models}}}\ (\bibinfo  {publisher} {University of Chicago press},\ \bibinfo {year} {1989})\BibitemShut {NoStop}%
\bibitem [{\citenamefont {Sargent}\ and\ \citenamefont {Stachurski}(2019)}]{sargent2019neumann}%
  \BibitemOpen
  \bibfield  {author} {\bibinfo {author} {\bibfnamefont {T.~J.}\ \bibnamefont {Sargent}}\ and\ \bibinfo {author} {\bibfnamefont {J.}~\bibnamefont {Stachurski}},\ }\href@noop {} {\bibfield  {journal} {\bibinfo  {journal} {Online Textbook}\ } (\bibinfo {year} {2019})}\BibitemShut {NoStop}%
\bibitem [{\citenamefont {Neumann}(1945)}]{neumann1945model}%
  \BibitemOpen
  \bibfield  {author} {\bibinfo {author} {\bibfnamefont {J.~v.}\ \bibnamefont {Neumann}},\ }\href@noop {} {\bibfield  {journal} {\bibinfo  {journal} {The Review of Economic Studies}\ }\textbf {\bibinfo {volume} {13}},\ \bibinfo {pages} {1} (\bibinfo {year} {1945})}\BibitemShut {NoStop}%
\bibitem [{\citenamefont {Thompson}\ and\ \citenamefont {Weil}(1971)}]{thompson1971neumann}%
  \BibitemOpen
  \bibfield  {author} {\bibinfo {author} {\bibfnamefont {G.~L.}\ \bibnamefont {Thompson}}\ and\ \bibinfo {author} {\bibfnamefont {R.~L.}\ \bibnamefont {Weil}},\ }in\ \href@noop {} {\emph {\bibinfo {booktitle} {Contributions to the Von Neumann Growth Model}}}\ (\bibinfo  {publisher} {Springer},\ \bibinfo {year} {1971})\ pp.\ \bibinfo {pages} {139--154}\BibitemShut {NoStop}%
\bibitem [{\citenamefont {Vassena}\ and\ \citenamefont {Stadler}(2024)}]{vassena2024unstable}%
  \BibitemOpen
  \bibfield  {author} {\bibinfo {author} {\bibfnamefont {N.}~\bibnamefont {Vassena}}\ and\ \bibinfo {author} {\bibfnamefont {P.~F.}\ \bibnamefont {Stadler}},\ }\href@noop {} {\bibfield  {journal} {\bibinfo  {journal} {Proceedings of the Royal Society A}\ }\textbf {\bibinfo {volume} {480}},\ \bibinfo {pages} {20230694} (\bibinfo {year} {2024})}\BibitemShut {NoStop}%
\end{thebibliography}%


\begin{thebibliography}{0}%
\makeatletter
\providecommand \@ifxundefined [1]{%
 \@ifx{#1\undefined}
}%
\providecommand \@ifnum [1]{%
 \ifnum #1\expandafter \@firstoftwo
 \else \expandafter \@secondoftwo
 \fi
}%
\providecommand \@ifx [1]{%
 \ifx #1\expandafter \@firstoftwo
 \else \expandafter \@secondoftwo
 \fi
}%
\providecommand \natexlab [1]{#1}%
\providecommand \enquote  [1]{``#1''}%
\providecommand \bibnamefont  [1]{#1}%
\providecommand \bibfnamefont [1]{#1}%
\providecommand \citenamefont [1]{#1}%
\providecommand \href@noop [0]{\@secondoftwo}%
\providecommand \href [0]{\begingroup \@sanitize@url \@href}%
\providecommand \@href[1]{\@@startlink{#1}\@@href}%
\providecommand \@@href[1]{\endgroup#1\@@endlink}%
\providecommand \@sanitize@url [0]{\catcode `\\12\catcode `\$12\catcode `\&12\catcode `\#12\catcode `\^12\catcode `\_12\catcode `\%12\relax}%
\providecommand \@@startlink[1]{}%
\providecommand \@@endlink[0]{}%
\providecommand \url  [0]{\begingroup\@sanitize@url \@url }%
\providecommand \@url [1]{\endgroup\@href {#1}{\urlprefix }}%
\providecommand \urlprefix  [0]{URL }%
\providecommand \Eprint [0]{\href }%
\providecommand \doibase [0]{https://doi.org/}%
\providecommand \selectlanguage [0]{\@gobble}%
\providecommand \bibinfo  [0]{\@secondoftwo}%
\providecommand \bibfield  [0]{\@secondoftwo}%
\providecommand \translation [1]{[#1]}%
\providecommand \BibitemOpen [0]{}%
\providecommand \bibitemStop [0]{}%
\providecommand \bibitemNoStop [0]{.\EOS\space}%
\providecommand \EOS [0]{\spacefactor3000\relax}%
\providecommand \BibitemShut  [1]{\csname bibitem#1\endcsname}%
\let\auto@bib@innerbib\@empty
\end{thebibliography}%

\putbib
\end{bibunit}

\end{document}